\documentclass[preprint,5p,twocolumn]{elsarticle}

\usepackage[english]{babel}
\usepackage{amsfonts,amsmath,amssymb,amsthm,bbold,bm,mathtools,slashed,empheq}
\usepackage{cancel,comment,enumerate,footnote,graphicx,subfloat,relsize}
\usepackage{array,tabularx,tabu,multirow,framed}
\usepackage[usenames,dvipsnames]{xcolor}
\usepackage[font=small]{caption}

\biboptions{sort&compress}
\usepackage{cleveref}
\crefformat{pluralequation}{#2\black{eqs.~(}#1\black{)}#3}
\Crefformat{pluralequation}{#2\black{Equations~(}#1\black{)}#3}
\crefformat{pluralequationfigure}{#2\black{figs.~}#1#3}
\Crefformat{pluralequationfigure}{#2\black{Figures~}#1#3}
\newcommand{\nc}{\newcommand}

\newcommand{\black}[1]	{{\color{black} 	#1}}

\nc{\nn}{\nonumber}
\nc{\hc}{\text{h.c.}}
\nc{\cc}{\text{c.c.}}

\def\[{\left[}
\def\]{\right]}
\def\({\left(}
\def\){\right)}
\def\<{\langle}
\def\>{\rangle}

\def\g5{\gamma_{5}}

\def\GeV{{\rm GeV}}
\def\TeV{{\rm TeV}}

\def \cm{{\rm cm}}

\def\ann		{{\rm ann}}
\def\Bcal		{\mathsmaller{\cal B}}
\def\BSF		{\mathsmaller{\rm BSF}}
\def\dec		{{\rm dec}}
\def\det		{{\rm det}}

\def\eff		{{\rm eff}}

\def\ion		{{\rm ion}}
\def\eq			{{\rm eq}}

\def\mpl 		{m_{\rm Pl}}

\def\one		{\mathsmaller{\mathbf{1}}}

\def\trans		{{\rm trans}}
\def\tot		{{\rm tot}}
\def\two		{\mathsmaller{\mathbf{2}}}

\def\vrel		{v_{\rm rel}}

\newcommand\widefbox[1]{\fbox{\hspace{1ex}#1\hspace{1ex}}}

\begin{document}
\begin{frontmatter}


\title{Saha equilibrium for metastable bound states and dark matter freeze-out}

\author[a]{Tobias Binder}
\ead{tobias.binder@ipmu.jp}

\author[b]{Anastasiia Filimonova}
\ead{a.filimonova@nikhef.nl}

\author[b,c]{Kalliopi Petraki\corref{CA}}
\ead{kalliopi.petraki@sorbonne-universite.fr}

\author[a]{Graham White}
\ead{graham.white@ipmu.jp}

\cortext[CA]{Corresponding author}

\address[a]{Kavli IPMU (WPI), UTIAS, The University of Tokyo, Kashiwa, Chiba 277-8583, Japan}
\address[b]{Nikhef, Science Park 105, 1098 XG Amsterdam, The Netherlands}
\address[c]{Sorbonne Universit\'e, CNRS, Laboratoire de Physique Th\'eorique et Hautes Energies (LPTHE), F-75252 Paris, France}


\begin{abstract}
The formation and decay of metastable bound states can significantly decrease the thermal-relic dark matter density, particularly for dark matter masses around and above the TeV scale. Incorporating bound-state effects in the dark matter thermal decoupling requires in principle a set of coupled Boltzmann equations for the bound and unbound species. However, decaying bound states attain and remain in a quasi-steady state. Here we prove in generality that this reduces the coupled system into a single Boltzmann equation of the standard form, with an effective cross-section that describes the interplay among bound-state formation, ionisation, transitions and decays. We derive a closed-form expression for the effective cross-section for an arbitrary number of bound states, and show that bound-to-bound transitions can only increase it. Excited bound levels may thus decrease the dark matter density more significantly than otherwise estimated. Our results generalise the Saha ionisation equilibrium to metastable bound states, potentially with applications beyond the dark matter thermal decoupling.
\end{abstract}


\end{frontmatter}


\section{Introduction \label{Sec:Intro}}

The thermal decoupling of dark matter (DM) from the primordial plasma is among the most widely considered mechanisms for its production. In thermal-relic DM scenarios, and for DM masses around and above the TeV scale, the interactions of DM or its co-annihilating partners typically manifest as long-range due to the hierarchy between the large DM mass and the smaller mass of the force mediators. Multi-TeV scenarios that do not feature such a hierarchy typically fail to provide for sufficient DM annihilation in the early universe and are thus not viable in this context, as suggested by model-independent unitarity arguments~\cite{Hisano:2002fk,Baldes:2017gzw}.

Long-range interactions imply the existence of bound states. The formation of \emph{meta-stable bound states} and their subsequent decays into radiation open new annihilation channels for DM that can significantly reduce its predicted relic density~\cite{vonHarling:2014kha}.  Besides being supported by unitarity arguments, the emergence of this type of inelasticity in the multi-TeV regime has been shown by explicit calculations in a variety of theories~\cite{
vonHarling:2014kha,
Ellis:2015vaa,
Baldes:2017gzw,
Harz:2018csl,
Harz:2019rro,
Harz:2017dlj,
Kim:2016zyy,
Biondini:2017ufr,
Biondini:2018pwp,
Biondini:2018ovz,
Biondini:2019int,
Binder:2018znk,
Ko:2019wxq,
Oncala:2019yvj,
Binder:2019erp,
Binder:2020efn,
Binder:2021otw,
Oncala:2021tkz,
Oncala:2021swy,
Bottaro:2021snn}. 
In addition to DM freeze-out, its implications extend to  DM indirect detection and collider signatures, as discussed in the original works~\cite{Pospelov:2008jd,MarchRussell:2008tu,Shepherd:2009sa}
and more recently in~\cite{An:2016gad,An:2016kie,Kouvaris:2016ltf,Asadi:2016ybp,Cirelli:2016rnw,Baldes:2017gzu,Cirelli:2018iax}.\footnote{
Stable bound states of typically asymmetric DM~\cite{Petraki:2013wwa} have an even longer history~\cite{Nussinov:1985xr,Kusenko:1997si,Foot:2004pa}, and more severe implications for DM phenomenology. These include affecting 
the DM self-interactions inside halos~\cite{Kusenko:2001vu,CyrRacine:2012fz,Petraki:2014uza,Wise:2014jva}, 
the DM direct detection signatures~\cite{Laha:2013gva,Cline:2012is,Kahlhoefer:2020rmg}, 
and giving rise to novel indirect detection signals~\cite{Pearce:2013ola,Pearce:2015zca,Cline:2014eaa,Detmold:2014qqa,Mahbubani:2019pij,Baldes:2020hwx}.}

If metastable bound states exist in the spectrum, then DM freeze-out in the early universe is governed by a system of coupled Boltzmann equations for the unbound species and the bound states~\cite{vonHarling:2014kha}. Besides the standard (co-)annihilations of the unbound particles directly into radiation~\cite{Griest:1990kh,Gondolo:1990dk}, this system  includes the bound-state formation (BSF), ionisation and decay processes, as well as bound-to-bound transitions. The latter have been frequently (albeit not always) neglected in the past, with most previous computations focusing on the effect of the ground states on the DM density~\cite{vonHarling:2014kha,Ellis:2015vaa,Baldes:2017gzw,Harz:2018csl,Harz:2019rro}. The reason has been two-fold: the capture into the ground state is indeed dominant in some models~\cite{Petraki:2015hla,Petraki:2016cnz,Oncala:2018bvl}, while excited states tend to decay into radiation more slowly and be ionised rapidly until lower temperatures than the ground state, when the DM density is smaller, which limits their efficacy in depleting the DM abundance. However, several studies have now shown that this does not hold in general; capture into excited states can be faster than into the ground state, while de-excitations increase (decrease) the effective decay (ionisation) rate of the excited states~\cite{Oncala:2019yvj,Oncala:2021tkz,Oncala:2021swy,Bottaro:2021snn}. Considering excited levels is hence crucial for predicting the DM density accurately~\cite{Oncala:2021swy,Bottaro:2021snn}. 

Clearly, a coupled system of equations for a large number of bound levels can be computationally demanding to solve. However, bound states that decay either directly into radiation or via transitions to other bound levels, attain and remain always in a quasi-steady state in an expanding universe~\cite{Ellis:2015vaa}. At high temperatures, this is ensured by the rapid BSF and ionisation processes, while at low temperatures the (effective) decay rate typically surpasses the expansion rate of the universe. As we shall show, the quasi-steady state that emerges depends on the interplay of the bound-state ionisation, transition and decay processes, and generalises the Saha ionisation equilibrium to metastable bound states. In this quasi-steady state, the equations that govern the DM freeze-out simplify significantly. More generally, this extension of Saha equilibrium pertains to all systems with metastable bound states in any thermodynamic environment.

Considering the above, in this work, we develop a general formalism for the DM thermal decoupling in the presence of an arbitrary number of metastable bound states, allowing for transitions among them. Under the assumption that the bound levels are in a quasi-steady state, we prove that the coupled system of equations reduces to a single Boltzmann equation that has the standard form of a Boltzmann equation for DM freeze-out with an attractor solution and an effective DM depletion cross-section. 
We show that the \emph{attractor solution} remains the standard equilibrium solution for the free species in the absence of bound states and derive the \emph{effective DM depletion cross-section in closed form}, in terms of the bound-state ionisation, transition and decay rates, \emph{for an arbitrary number of bound states}. This single object encapsulates the system's entire dynamics, and provides a simple way to incorporate BSF in DM freeze-out computations and public DM codes.

The same object also specifies how Saha equilibrium varies between the two known regimes of stable and rapidly decaying bound states, and allows insight into the effect of bound-to-bound transitions. We prove that, despite the complex interplay of the various processes involved,   
\emph{bound-to-bound transitions can only increase the effective DM depletion rate}.
This is an important result that renders excited bound levels more important than previously anticipated.
In addition, in the limit of very rapid transitions, we show that the effective DM depletion rate becomes independent of the bound-to-bound transition rates.

We note that our formalism is independent of the microphysics that underlies the relevant processes. Depending on the model, bound states may form, get ionised and transition between each other radiatively~\cite{vonHarling:2014kha,Ellis:2015vaa,Petraki:2015hla,Petraki:2016cnz,Baldes:2017gzw,Harz:2018csl,Harz:2019rro,Oncala:2019yvj,Ko:2019wxq,Oncala:2021tkz,Oncala:2021swy,Bottaro:2021snn}, via scattering on the relativistic plasma~\cite{Binder:2019erp,Binder:2020efn,Binder:2021otw,Oncala:2021tkz,Oncala:2021swy} or other re-arrangement processes~\cite{Geller:2018biy,Geller:2020zhq}. Our parametrisation encompasses all possibilities, and our analysis pertains to all scenarios that feature metastable bound states.

This paper is organised as follows. In \cref{Sec:BoltzmannEqs}, we introduce the coupled system of Boltzmann equations that describe the DM freeze-out in the presence of bound states, reduce it to a single equation with an effective DM depletion cross-section that we derive in closed form, and discuss the generalised Saha ionisation equilibrium that emerges. In \cref{Sec:Transitions}, we examine the effect of bound-to-bound transitions and show that they can only increase the effective DM depletion cross-section. We illustrate our results in a toy model in \cref{Sec:ToyModel}, and conclude in \cref{Sec:Concl}.

\section{Boltzmann equations \label{Sec:BoltzmannEqs}}

\subsection{Coupled system \label{sec:BoltzmannEqs_Coupled}}

For generality, in the following, we consider the possibility of several co-annihilating species that share the same quantum number that stabilises DM. Bound states can form from any combination of these species. Let $Y_j \equiv n_j/s$ and  $Y_{\Bcal} \equiv n_{\Bcal}/s$ be the number-density-to-entropy-density ratios of the free species $j$ and the bound state ${\cal B}$ respectively. As is standard, we use the time parameter
\begin{align}
x \equiv m/T ,
\label{eq:x_def}
\end{align}
where $m$ can be taken to be the mass of the lightest co-annihilating species. 
The energy and entropy densities of the universe are
$\rho= (\pi^2/30) g_{*\rho} T^4$ and 
$s= (2\pi^2/45) g_{*\mathsmaller{S}} T^3$, 
where $g_{*\rho}$ and $g_{*\mathsmaller{S}}$ stand for the corresponding degrees of freedom (dof) of the universe respectively. We define~\cite{Gondolo:1990dk} 
\begin{align}
g_{*}^{1/2} &= 
\frac{g_{*\mathsmaller{S}}}{\sqrt{g_{*\rho}}}
\(1-\frac{x}{3g_{*\mathsmaller{S}}} \frac{dg_{*\mathsmaller{S}}}{dx} \) .
\label{eq:gstareff}
\end{align}
The evolution of $Y_j$ and $Y_{\Bcal}$ is governed by the coupled Boltzmann equations
\begin{subequations}
\label{eq:BoltzmannEqs}
\label[pluralequation]{eqs:BoltzmannEqs}
\begin{align}
\frac{dY_j}{dx} =
&- \frac{\lambda}{x^2} \sum_i \<\sigma_{ji}^{\ann} \vrel\> \(Y_j Y_i - Y_j^{\eq}Y_i^{\eq}\)  
\nn \\ 
&- \frac{\lambda}{x^2} \sum_i \sum_{\Bcal}\<\sigma_{j i\to \Bcal}^{\BSF} \, \vrel\> 
\(Y_j Y_i - \frac{Y_\Bcal}{Y_{\Bcal}^{\eq}} Y^{\eq}_j Y^{\eq}_i\) 
\nn \\  
&- \Lambda \, x  \sum_{i} \<\Gamma_{j\to i}\> 
\(Y_j - \dfrac{Y_i}{Y_i^{\eq}} Y_j^{\eq}\) ,
\label{eq:BoltzmannEqs_Free} 
\\
\frac{dY_\Bcal}{dx} =&
- \Lambda \, x \[ 
\< \Gamma_{\Bcal}^{\dec} \>   \(Y_{\Bcal} - Y_{\Bcal}^{\eq} \)
\right. \nn \\ &\left. 
+\sum_{i,j} \< \Gamma_{\Bcal \to ij}^{\ion} \>
\(Y_{\Bcal} -  \dfrac{Y_i Y_j}{Y_i^{\eq} Y_j^{\eq}} Y_{\Bcal}^{\eq}\)
\right. \nn \\ &\left. 
+ \sum_{\Bcal' \neq \Bcal} 
\<\Gamma_{\Bcal\to \Bcal'}^{\trans} \>
\(Y_{\Bcal} - \dfrac{Y_{\Bcal'}}{Y_{\Bcal'}^{\eq}} Y_{\Bcal}^{\eq} \)
\] ,
\label{eq:BoltzmannEqs_Bound} 
\end{align}
\end{subequations}
where 
\begin{subequations}
\label{eq:lambdaAndLambda}
\label[pluralequation]{eqs:lambdaAndLambda}
\begin{align}
\lambda &\equiv 
\sqrt{\frac{\pi}{45}}\mpl \, m \, g_{*}^{1/2} ,
\label{eq:lambda}
\\
\Lambda &\equiv \dfrac{\lambda}{s \, x^3} 
=\sqrt{\frac{45}{4\pi^3}} \frac{\mpl}{m^2} \frac{g_{*}^{1/2}}{g_{*\mathsmaller{S}}} .
\label{eq:Lambda}
\end{align}
\end{subequations}
The equilibrium densities in the non-relativistic regime are
\begin{subequations}
\label{eq:EquilibriumDensities}
\label[pluralequation]{eqs:EquilibriumDensities}	
\begin{align}
Y_j^\eq &\simeq \frac{90}{(2\pi)^{7/2}} 
\frac{g_{i,\eff}}{g_{*\mathsmaller{S}}} 
\, x^{3/2} \, e^{-x} ,
\label{eq:Yeq_i}
\\
Y_{\Bcal}^\eq &\simeq \frac{90}{(2\pi)^{7/2}} 
\frac{g_{\Bcal,\eff}}{g_{*\mathsmaller{S}}} 
\, (2x)^{3/2} 
\, e^{-2x} \, e^{|{\cal E}_{\Bcal}|/T} ,
\label{eq:Yeq_B}
\end{align}
\end{subequations}
where ${\cal E}_{\Bcal}<0$ is the binding energy of the ${\cal B}$ bound state.
The effective dof depend on the temperature as follows
\begin{subequations}
\label{eq:dof_effective}
\label[pluralequation]{eqs:dof_effective}
\begin{align}
g_{i,\eff} &\equiv g_i (1+\Delta_i)^{3/2} e^{-x\Delta_i} ,
\label{eq:dof_effective_FreeSpecies} 
\\
g_{\Bcal,\eff} &\equiv g_{\Bcal} [1+\Delta_{\Bcal} - |{\cal E}_{\Bcal}|/(2m)]^{3/2} 
e^{-2x \, \Delta_{\Bcal}} ,
\label{eq:dof_effective_BoundStates}
\end{align}
\end{subequations}
with $g_i$ and $g_{\Bcal}$ being the actual dof of the free species and bound states. We defined the fractional mass differences
\begin{subequations}
\label{eq:MassDifferencesFractional}
\label[pluralequation]{eqs:MassDifferencesFractional}
\begin{align}
\Delta_j &\equiv (m_j - m)/m , 
\label{eq:MassDifferencesFractional_Free}
\\
\Delta_{\Bcal} &\equiv (m_{\Bcal} + |{\cal E}_{\Bcal}| - 2m)/(2m) .
\label{eq:MassDifferencesFractional_Bound}
\end{align}
\end{subequations}
We are ultimately interested in the total yield of all co-annihilating species
\begin{align}
Y \equiv \sum_j  Y_j ,
\label{eq:DensityTotal_def}
\end{align}
whose equilibrium value is 
\begin{align}
Y^\eq &= \frac{90}{(2\pi)^{7/2}} 
\frac{g_{\eff}}{g_{*\mathsmaller{S}}} 
\, x^{3/2} \, e^{-x} ,
\label{eq:Yeq}
\end{align}
with
\begin{align}
g_{\eff} 
\equiv \sum_j g_{j,\eff} 
= \sum_j g_j (1+\Delta_j)^{3/2} e^{-x \Delta_j} .
\label{eq:gDM_def}
\end{align}
Clearly, $Y_j^{\eq} = (g_{j,\eff}/g_{\eff}) Y^{\eq}$. Note that the bound states are metastable and their abundance becomes eventually negligible, so we do not include them in \cref{eq:DensityTotal_def}.

In \cref{eqs:BoltzmannEqs}, $\Gamma_{\Bcal}^{\dec}$, $\Gamma_{\Bcal\to ij}^{\ion}$ and $\Gamma_{\Bcal\to\Bcal'}^{\trans}$ are, respectively, the rates of ${\cal B}$ decay into radiation, ${\cal B}$ ionisation (a.k.a.~dissociation) into $ij$, and ${\cal B}$ transition into another bound level ${\cal B}'$. The rates $\Gamma_{j\to i}$ describe the transitions between free particles, due to decays, inverse decays and/or scatterings on the thermal bath; overall, these processes do not change the DM number density, but retain kinetic and chemical equilibrium among the dark species. 
Note that in \cref{eqs:BoltzmannEqs}, we must use the thermally averaged rates, $\<\Gamma\>$. The thermal average introduces Lorentz dilation factors for decay processes, which however are insignificant in the non-relativistic regime, as well as Bose-enhancement factors in the case of bound-to-bound transitions and capture or ionisation processes. The thermally-averaged rates and cross-sections of inverse processes are related via detailed balance that we have already employed in writing \cref{eqs:BoltzmannEqs}, 
\begin{subequations}
\label{eq:DetailedBalance}
\label[pluralequation]{eqs:DetailedBalance}
\begin{align}
\<\Gamma_{\Bcal \to ij}^{\ion}\> &= s \,
\<\sigma_{ij \to \Bcal}^{\BSF} \vrel \> \times
(Y_i^{\rm eq}Y_j^{\rm eq} / Y_{\Bcal}^{\rm eq}) ,
\label{eq:DetailedBalance_IonBSF}
\\
\<\Gamma_{\Bcal \to \Bcal'}^{\trans}\> &= 
\<\Gamma_{\Bcal'\to \Bcal}^{\trans} \> \times
(Y_{\Bcal'}^{\rm eq} / Y_{\Bcal}^{\rm eq}) ,
\label{eq:DetailedBalance_B2Btransitions}
\\
\<\Gamma_{i\to j}\> &= 
\<\Gamma_{j\to i}\> \times (Y_j^{\rm eq} / Y_i^{\rm eq}) .
\label{eq:DetailedBalance_FreeSpeciesScattering} 
\end{align}
\end{subequations}

Typically the various co-annihilating species either down-scatter or decay into the lightest one, that then makes up all of the DM today. Assuming that $m$ has been chosen to be its mass, the fractional DM relic density is\footnote{
Even if several co-annihilating species retain significant relic abundance, \cref{eq:OmegaDM} is usually a good approximation, since the mass difference among co-annihilating species typically should not exceed 10\% in order for co-annihilations to be efficient at temperatures $T \lesssim m/30$. Larger mass differences necessitate following the chemical decoupling of each species separately in order to predict the DM density precisely; however they also render down-scattering and decays more efficient, making it less likely that the heavier co-annihilating partners have a significant relic density.}
%
\begin{align}
\Omega \simeq f \, m \, Y_\infty s_0 / \rho_c, 
\label{eq:OmegaDM}
\end{align} 
where $Y_\infty$ is the final value of the total yield \eqref{eq:DensityTotal_def} and $f = 1$~or~2 if the lightest species is self-conjugate or non-self-conjugate respectively.\footnote{If the lightest species is self-conjugate and there are non-self-conjugate co-annihilating partners, then in \cref{eq:DensityTotal_def} we must include the dof of both particles and antiparticles of those partners.} 
In \cref{eq:OmegaDM}, $s_0 \simeq 2839.5~\cm^{-3}$ and $\rho_c \simeq 4.78 \cdot 10^{-6}~\GeV~\cm^{-3}$ are the entropy density and critical energy density of the universe today~\cite{Aghanim:2018eyx}.

\subsection{Effective equation \label{sec:BoltzmannEqs_Effective}}

The system of Boltzmann \cref{eqs:BoltzmannEqs} is computationally expensive to solve. However, it is possible under reasonable assumptions that remain valid throughout the DM chemical decoupling from the primordial plasma, to reduce it to a single effective Boltzmann equation, as we shall now see. In particular: 
\begin{enumerate}[(i)]
\item 	
We assume that the $i \leftrightarrow j$ conversion processes remain sufficiently fast to keep the various free species in chemical equilibrium; this implies
\begin{subequations}
\label{eq:BoltzEqsEff_Assumptions}
\label[pluralequation]{eqs:BoltzEqsEff_Assumptions}
\begin{align}
Y_i / Y_i^\eq = w, 
\label{eq:BoltzEqsEff_Assumptions_ChemicalEquil}
\end{align}
where $w$ is the same for all $i$, and in general depends on $x$; it essentially reparametrises the chemical potential of the free species, $w=\exp(\mu_{\rm free}/T)$. The assumption \eqref{eq:BoltzEqsEff_Assumptions_ChemicalEquil} appeared first in the context of DM in ref.~\cite{Griest:1990kh}. It is satisfied in many particle physics models; however, if species carrying the quantum number that stabilises DM are not in chemical equilibrium with the latter, then they have to be treated separately (see, e.g., \cite{DAgnolo:2017dbv,Garny:2017rxs}).

\item 
At high temperatures, BSF and ionization processes are very efficient, while later on decays into radiation, directly or via bound-to-bound transitions, dominate. In both cases, the rates that appear in the right-hand side of \cref{eq:BoltzmannEqs_Bound} remain very large such that the bound-state densities attain a quasi-steady state where
\begin{align}
dY_{\Bcal} / dx \simeq 0. 
\label{eq:BoltzEqsEff_Assumptions_BoundStateDerivative0}
\end{align}
This assumption was first proposed in ref.~\cite{Ellis:2015vaa}, and yields a system of linear equations for $Y_{\Bcal}$ whose solution can be re-employed in \cref{eq:BoltzmannEqs_Free}. 
We will discuss the validity of this approximation in \cref{sec:BoltzmannEqs_ValidityApprox}, using the solution that we obtain in the following.

\end{subequations}
\end{enumerate}

Under the assumption \eqref{eq:BoltzEqsEff_Assumptions_ChemicalEquil}, the Boltzmann eqs.~\eqref{eq:BoltzmannEqs_Free} imply that the total yield \eqref{eq:DensityTotal_def} obeys the equation
\begin{align}
\dfrac{dY}{dx} &= -\dfrac{\lambda}{x^2} \sum_{i,j} Y_i^{\eq}Y_j^{\eq}  
\[ 
\<\sigma_{ij}^{\ann} \vrel\> (w^2-1) 
\right. \nn \\ &\left.
+ \sum_{\Bcal} \<\sigma_{ij \to \Bcal}^{\BSF} \, \vrel\> 
\(w^2- Y_\Bcal / Y_{\Bcal}^{\eq}\)
\],
\label{eq:BoltzmannEqs_Ytotal} 
\end{align}
where we note that the $i \leftrightarrow j$ conversion terms have canceled each other, as they do not change the total density.
Using the definitions of the previous section, it is straightforward to show that 
\begin{align}
Y_i^\eq Y_j^\eq (w^2-1) = 
\dfrac{g_{i,\eff} g_{j,\eff}}{g_{\eff}^2} \[Y^2 - (Y^{\eq})^2\] ,    
\label{eq:w2-1}
\end{align} 
and thus bring the first term of \cref{eq:BoltzmannEqs_Ytotal} in the canonical form, with $Y^{\eq}$ being the attractor solution and 
$\sum_{i,j} (g_{i,\eff} g_{j,\eff} / g_{\eff}^2) \< \sigma^{\ann}_{ij} \vrel \>$
being the effective annihilation cross-section~\cite{Griest:1990kh}.

Under the assumption \eqref{eq:BoltzEqsEff_Assumptions_BoundStateDerivative0}, it is also straightforward to express the bound-state term of \cref{eq:BoltzmannEqs_Ytotal} in terms of the actual densities of the free species only and the equilibrium densities of both the free and bound states. However, the attractor solution and the form of the effective cross-section that arise from this term are not immediately evident. 
Using both the assumption \eqref{eq:BoltzEqsEff_Assumptions_BoundStateDerivative0} and detailed balance, in the following we prove that the attractor solution for $Y$ remains $Y^\eq$ and determine the bound-state contribution to the effective cross-section in closed form.

We begin with defining the total ionisation and bound-to-bound transition rates, as well as the total interaction rate for the bound level ${\cal B}$,
\begin{subequations}
\label{eq:Rates_IonTrans_def}	
\label[pluralequation]{eqs:Rates_IonTrans_def}	
\begin{align}
\< \Gamma_{\Bcal}^\ion \> &\equiv 
\sum_{i,j}  \< \Gamma_{\Bcal \to ij}^\ion \>,
\label{eq:Rate_Ion_def}	
\\
\< \Gamma_{\Bcal}^\trans \> &\equiv 
\sum_{\Bcal' \neq \Bcal}  
\< \Gamma_{\Bcal \to \Bcal'}^\trans \> ,
\label{eq:Rate_Trans_def}	
\\
\< \Gamma_{\Bcal}^\tot \> &\equiv   
\< \Gamma_{\Bcal}^\dec \> +
\< \Gamma_{\Bcal}^\ion \> +
\< \Gamma_{\Bcal}^\trans \> ,
\label{eq:Rate_Tot_def}	
\end{align}
\Cref{eq:Rate_Ion_def,eq:DetailedBalance_IonBSF} allow to define the total capture cross-section into a level ${\cal B}$,
\begin{align}
\<\sigma_{\Bcal}^{\BSF} \vrel \> \equiv 
\<\Gamma_{\Bcal}^{\ion}\> \times \dfrac{Y_{\Bcal}^{\eq} }{s (Y^{\eq})^2} =
\sum_{i,j} \dfrac{g_{i,\eff} g_{j,\eff}}{g_{\eff}^2}    
\<\sigma_{ij\to\Bcal}^{\BSF} \vrel \> .
\label{eq:sigmaBSF_tot_def}	
\end{align}
\end{subequations}
Then, we define the column vectors of the bound-state yields $\mathbb{Y}$ and $\mathbb{Y}^\eq$,\footnote{$\mathbb{Y}$ and $\mathbb{Y}^\eq$ should not be confused with the yields $Y$ and $Y^\eq$ of the unbound species; we never use matrix notation for the latter in the present work.} 
\begin{align}
\mathbb{Y}_{\Bcal} = Y_{\Bcal},  \qquad
\mathbb{Y}_{\Bcal}^\eq = Y_{\Bcal}^{\eq}, 
\label{eq:Matrices_Yields}
\end{align}
as well as the square matrices 
$\mathbb{\Gamma}^\dec$, 
$\mathbb{\Gamma}^\ion$, 
$\mathbb{\Gamma}^\trans$, 
$\mathbb{\Gamma}^\tot$, 
$\mathbb{T}$,
\begin{subequations}
\label{eq:Matrices_Rates}
\label[pluralequation]{eqs:Matrices_Rates}	
\begin{align}
\mathbb{\Gamma}^\dec_{\Bcal \Bcal'} 
&\equiv \delta_{\Bcal \Bcal'} \< \Gamma^{\dec}_{\Bcal} \>,
\label{eq:Matrices_GammaDec}
\\
\mathbb{\Gamma}^\ion_{\Bcal \Bcal'} 
&\equiv \delta_{\Bcal \Bcal'} \< \Gamma^{\ion}_{\Bcal} \>,
\label{eq:Matrices_GammaIon}
\\
\mathbb{\Gamma}^\trans_{\Bcal \Bcal'} 
&\equiv \delta_{\Bcal \Bcal'} \< \Gamma^{\trans}_{\Bcal} \>,
\label{eq:Matrices_GammaTrans}
\\
\mathbb{\Gamma}^\tot_{\Bcal \Bcal'} 
&\equiv \delta_{\Bcal \Bcal'}  \< \Gamma^{\tot}_{\Bcal} \>
= \mathbb{\Gamma}^{\dec}_{\Bcal \Bcal'}
+ \mathbb{\Gamma}^{\ion}_{\Bcal \Bcal'} 
+ \mathbb{\Gamma}^{\trans}_{\Bcal \Bcal'}  ,
\label{eq:Matrices_GammaTot}
\\
\mathbb{T}_{\Bcal \Bcal'} 
&\equiv \< \Gamma^{\trans}_{\Bcal' \to \Bcal} \> .
\label{eq:Matrices_T}
\end{align}
\end{subequations}
We note the opposite order of the indices on the left and right sides of \cref{eq:Matrices_T}. No summation over repeated indices is implied in any of \cref{eqs:Matrices_Rates}. Clearly,  
$\mathbb{\Gamma}^\dec$, 
$\mathbb{\Gamma}^\ion$, 
$\mathbb{\Gamma}^\trans$ and 
$\mathbb{\Gamma}^\tot$ 
are diagonal, while  
$\mathbb{T}$ has only off-diagonal non-vanishing elements.

Using this matrix notation and under the assumption \eqref{eq:BoltzEqsEff_Assumptions_BoundStateDerivative0}, the Boltzmann \cref{eq:BoltzmannEqs_Bound} for the bound species yields the linear system 
\begin{align} 
\(\mathbb{\Gamma}^\tot -\mathbb{T} \) \mathbb{Y} =  
\(\mathbb{\Gamma}^\dec + w^2\mathbb{\Gamma}^\ion \) \mathbb{Y}^{\eq}.
\label{eq:YBound}
\end{align}
We must now solve \cref{eq:YBound} for $\mathbb{Y}$. First, we consider the detailed balance \cref{eq:DetailedBalance_B2Btransitions} for bound-to-bound transitions, which in matrix form reads
\begin{subequations}
\label{eq:DetailedBalance_MatrixForms}
\label[pluralequation]{eqs:DetailedBalance_MatrixForms}    
\begin{align}
\mathbb{T}_{\Bcal\Bcal'} \mathbb{Y}_{\Bcal'}^{\eq} = 
\mathbb{T}_{\Bcal'\Bcal} \mathbb{Y}_{\Bcal}^{\eq} .
\label{eq:DetailedBalance_Tmatrix}    
\end{align}
Summing over ${\cal B}'$,
\begin{align}
\mathbb{\Gamma}^{\trans} \, \mathbb{Y}^{\eq} = 
\mathbb{T} \, \mathbb{Y}^\eq , 
\label{eq:DetailedBalance_GammaY=TY}
\end{align}
or equivalently
\begin{align}
\(\mathbb{\Gamma}^{\tot} - \mathbb{T}\) \mathbb{Y}^\eq
= \(\mathbb{\Gamma}^{\dec} + \mathbb{\Gamma}^{\ion} \) \mathbb{Y}^\eq .
\label{eq:DetailedBalance_GammaY=TY_modified}
\end{align}  
We also note that since $\mathbb{\Gamma}^{\tot}$ is diagonal, $\mathbb{\Gamma}^{\tot}-\mathbb{T}$ and its inverse also possess the property \eqref{eq:DetailedBalance_Tmatrix}, namely 
\begin{align}
(\mathbb{\Gamma}^{\tot}-\mathbb{T})_{\Bcal \Bcal'} \mathbb{Y}_{\Bcal'}^\eq &=
(\mathbb{\Gamma}^{\tot}-\mathbb{T})_{\Bcal' \Bcal}
\mathbb{Y}_{\Bcal}^\eq ,
\label{eq:DetailedBalance_GammaTotMinusT}
\\
(\mathbb{\Gamma}^{\tot}-\mathbb{T})^{-1}_{\Bcal \Bcal'} \mathbb{Y}_{\Bcal'}^\eq &=
(\mathbb{\Gamma}^{\tot}-\mathbb{T})^{-1}_{\Bcal' \Bcal} \mathbb{Y}_{\Bcal}^\eq .
\label{eq:DetailedBalance_GammaTotMinusT_Inverse}
\end{align}
\end{subequations}
\Cref{eq:DetailedBalance_GammaTotMinusT} is evident. Then, it is easy to check that if 
$\mathbb{P}_{\Bcal \Bcal'} \equiv (\mathbb{\Gamma}^{\tot}-\mathbb{T})^{-1}_{\Bcal \Bcal'}$ 
and 
$\tilde{\mathbb{P}}_{\Bcal \Bcal'} \equiv 
\mathbb{P}_{\Bcal' \Bcal} (Y_{\Bcal}^{\eq}/ Y_{\Bcal'}^{\eq})$, 
then
$\tilde{\mathbb{P}}(\mathbb{\Gamma}^{\tot}-\mathbb{T}) = \mathbb{1}$. 
By virtue of the uniqueness of the inverse, $\mathbb{P} = \tilde{\mathbb{P}}$, and \cref{eq:DetailedBalance_GammaTotMinusT_Inverse} follows.

Combining \cref{eq:YBound,eq:DetailedBalance_GammaY=TY_modified} to eliminate $\mathbb{\Gamma}^{\ion} \mathbb{Y}^{\eq}$, and solving for $\mathbb{Y}$, we find
\begin{align} 
\mathbb{Y} = w^2 \mathbb{Y}^{\eq}
+ (1-w^2) \(\mathbb{\Gamma}^\tot -\mathbb{T} \)^{-1}
\mathbb{\Gamma}^\dec  \mathbb{Y}^{\eq} .
\label{eq:YBound_mod}
\end{align}
Considering the ${\cal B}$ element of \cref{eq:YBound_mod}, we obtain
\begin{align} 
w^2 - \dfrac{Y_{\Bcal}}{Y_{\Bcal}^{\eq}} 
&=  (w^2-1) \dfrac{1}{Y^{\eq}_{\Bcal}}
\[\(\mathbb{\Gamma}^\tot -\mathbb{T} \)^{-1}
\mathbb{\Gamma}^\dec \mathbb{Y}^{\eq} \]_{\Bcal}
\nn \\
&=  (w^2-1) \sum_{\Bcal'} 
\<\Gamma_{\Bcal'}^\dec\> 
\(\mathbb{\Gamma}^\tot -\mathbb{T} \)^{-1}_{\Bcal'\Bcal} ,
\label{eq:BSF_ContributionToBoltzmann}
\end{align}
where in the second step we took into account \cref{eq:DetailedBalance_GammaTotMinusT_Inverse} and that $\mathbb{\Gamma}^{\dec}$ is diagonal. \Cref{eq:BSF_ContributionToBoltzmann} can be employed directly into the Boltzmann \cref{eq:BoltzmannEqs_Ytotal}, which with the help of \eqref{eq:w2-1}, takes now the canonical form
\begin{align}
\frac{dY}{dx} =& 
- \frac{\lambda}{x^2} \<\sigma^{\eff} \vrel\> \[ Y^2 - (Y^{\eq})^2 \]  ,
\label{eq:BoltzmannEqs_Eff} 
\end{align}
with the effective cross-section
\begin{align}
\<\sigma^{\eff} \vrel\> \equiv 
\sum_{i,j}  \! \dfrac{ g_{i,\eff} \, g_{j,\eff} }{g_{\eff}^2} \!
\[\<\sigma_{ij}^{\ann} \vrel\>  \!+\!
\sum_{\Bcal} r_{\Bcal} \, \<\sigma_{ij\to \Bcal}^{\BSF} \vrel\> \],
\label{eq:CrossSection_Effective}
\end{align}
where $g_{i,\eff}$ and $g_\eff$ are given by eqs.~\eqref{eq:dof_effective_FreeSpecies} and \eqref{eq:gDM_def}, and we define the \emph{efficiency factors}
\begin{empheq}[box=\widefbox]{align}
r_{\Bcal} \equiv 
\sum_{\Bcal'}
\<\Gamma^\dec_{\Bcal'}\>
\(\mathbb{\Gamma}^\tot -\mathbb{T} \)^{-1}_{\Bcal'\Bcal}.
\label{eq:EfficiencyFactors}
\end{empheq}

\Cref{eq:EfficiencyFactors}, in the context of \cref{eq:BoltzmannEqs_Eff,eq:CrossSection_Effective}, is one of the main results of the present work. It furnishes a general proof that the system of coupled Boltzmann \cref{eqs:BoltzmannEqs} can be reduced into a single one, under the assumptions \eqref{eqs:BoltzEqsEff_Assumptions}; this had previously been shown only in specific cases with a small number of bound states, while neglecting~\cite{Ellis:2015vaa} or considering~\cite{Oncala:2021swy,Bottaro:2021snn} bound-to-bound transitions. The closed form of \cref{eq:EfficiencyFactors} renders it suitable for incorporating a large number of bound levels to thermal decoupling calculations.

The efficiency factors are bounded by 
\begin{align}
0 \leqslant r_{\Bcal} \leqslant 1  ,
\label{eq:rB_limits}
\end{align}
where the limiting values are discussed in more detail in section~\ref{sec:BoltzmannEqs_Saha}. 
However, the inequality \eqref{eq:rB_limits} is not immediately evident from the definition \eqref{eq:EfficiencyFactors}. 
To prove it, we first note that \cref{eq:DetailedBalance_GammaY=TY_modified}, expressed as
$\sum_{\Bcal'} \(\mathbb{\Gamma}^\tot -\mathbb{T} \)^{-1}_{\Bcal \Bcal'}
\( \<\Gamma^\dec_{\Bcal'}\> + \<\Gamma^\ion_{\Bcal'}\> \) Y_{\Bcal'}^{\eq} 
=Y^{\eq}_{\Bcal}$,
together with \cref{eq:DetailedBalance_GammaTotMinusT_Inverse}, imply
\begin{align}
\sum_{\Bcal'} \<\Gamma^{\ion}_{\Bcal'}\>
\(\mathbb{\Gamma}^\tot -\mathbb{T} \)^{-1}_{\Bcal'\Bcal} 
&= 1- \sum_{\Bcal'} \<\Gamma^{\dec}_{\Bcal'}\>
\(\mathbb{\Gamma}^\tot -\mathbb{T} \)^{-1}_{\Bcal'\Bcal} 
\nn \\
&=  1-r_{\Bcal} .   
\label{eq:1-rB}
\end{align}
Then, considering \cref{eq:EfficiencyFactors,eq:1-rB}, it suffices to show that 
$\(\mathbb{\Gamma}^\tot -\mathbb{T} \)^{-1} \geqslant 0$ in order for \eqref{eq:rB_limits} to be true. 
This is so because $\mathbb{\Gamma}^\tot -\mathbb{T}$ is a (non-singular) $M$-matrix, since 
(i)~$\mathbb{\Gamma}^\tot -\mathbb{T}$ is a $Z$-matrix, i.e.~all its off-diagonal elements, $-\mathbb{T}$, are less than or equal to zero, and 
(ii)~there exists a vector $v > 0$ such that 
$\(\mathbb{\Gamma}^\tot -\mathbb{T} \) v > 0$; indeed, $v=\mathbb{Y}^\eq$ satisfies this requirement as per \cref{eq:DetailedBalance_GammaY=TY_modified}~\cite[theorem 1, condition $K_{33}$]{PLEMMONS1977175}.

\subsection{Generalisation of Saha ionisation equilibrium to metastable bound states\label{sec:BoltzmannEqs_Saha}}

To obtain insight in the dynamics of systems with metastable bound states, we now examine more carefully the evolution of the bound-state densities, using the results obtained in the above. This will ultimately also allow us in the following subsection to determine the range of validity of our approximation.
\Cref{eq:BSF_ContributionToBoltzmann} can be re-expressed using \eqref{eq:EfficiencyFactors}, as 
\begin{empheq}[box=\widefbox]{align} 
\dfrac{n_{\Bcal}}{n_{\Bcal}^{\eq}} = 
\(\dfrac{n_{\rm free}}{n_{\rm free}^{\eq}} \)^2
-\[\(\dfrac{n_{\rm free}}{n_{\rm free}^{\eq}} \)^2 -1\] r_{\Bcal},
\label{eq:SahaEquilibrium_Metastable}
\end{empheq}
where for generality we used the number densities $n = sY$, and set $w = n_{\rm free} / n_{\rm free}^{\eq} = \exp(\mu_{\rm free}/T)$, with `free' denoting any of the unbound co-annihilating species, presumed in chemical equilibrium with each other [cf.~\cref{eq:BoltzEqsEff_Assumptions_ChemicalEquil}]. 
Using \cref{eq:sigmaBSF_tot_def,eq:1-rB},  \cref{eq:SahaEquilibrium_Metastable} can be also expressed in the following form that will be useful below,
\begin{align}
n_{\Bcal} = r_{\Bcal} n_{\Bcal}^{\eq} 
+ n_{\rm free}^2  \sum_{\Bcal'}   
(\mathbb{\Gamma}^\tot - \mathbb{T})^{-1}_{\Bcal'\Bcal} 
\< \sigma_{\Bcal'}^{\BSF} \vrel \>
\frac{n_{\Bcal}^{\eq}}{n_{\Bcal'}^{\eq}} .
\tag{\ref{eq:SahaEquilibrium_Metastable}a}
\label{eq:SahaEquilibrium_Metastable_Alt}
\end{align} \Cref{eq:SahaEquilibrium_Metastable}, together with the expression~\eqref{eq:EfficiencyFactors} for $r_{\Bcal}$ in terms of the inherent parameters of the theory, generalise Saha equilibrium to metastable bound states. In terms of chemical potentials, it reads
\begin{align}
\mu_{\Bcal} / T = 2\mu_{\rm free}/T + \ln \[1 - (1-e^{-2\mu_{\rm free}/T}) r_{\Bcal} \].
\label{eq:SahaEquilibrium_Metastable_ChemicalPot}
\end{align}

\begin{figure}[t!]
\centering
\includegraphics[width=0.34\textwidth]{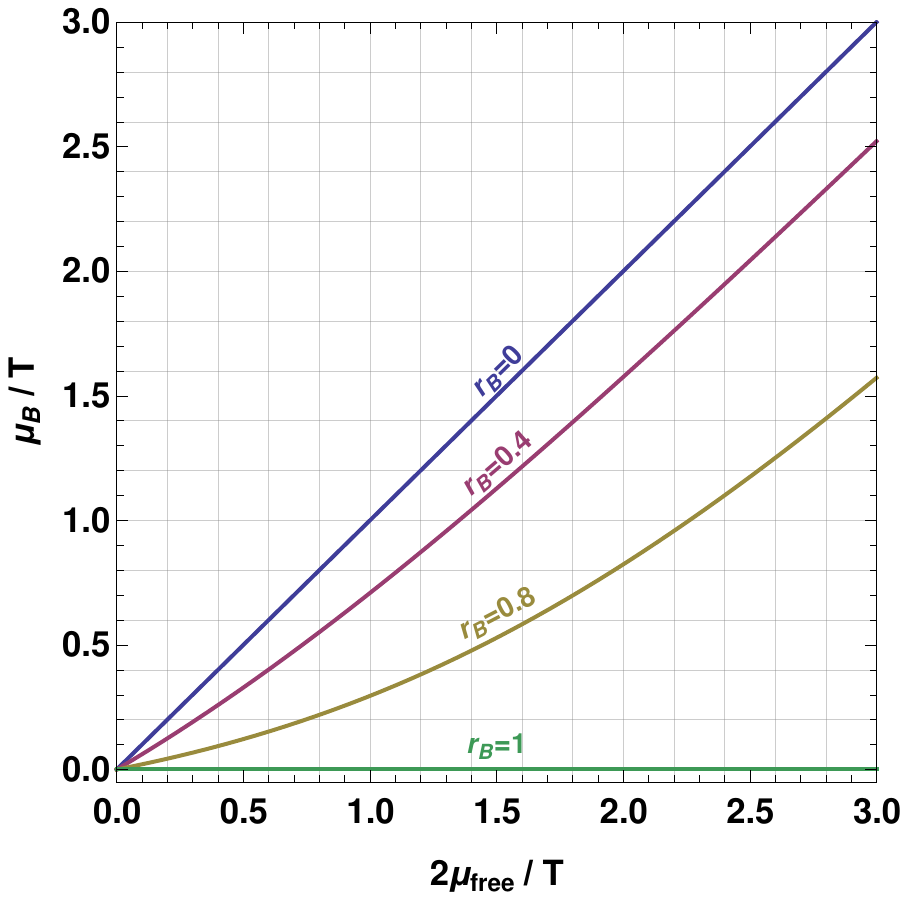}
\includegraphics[width=0.34\textwidth]{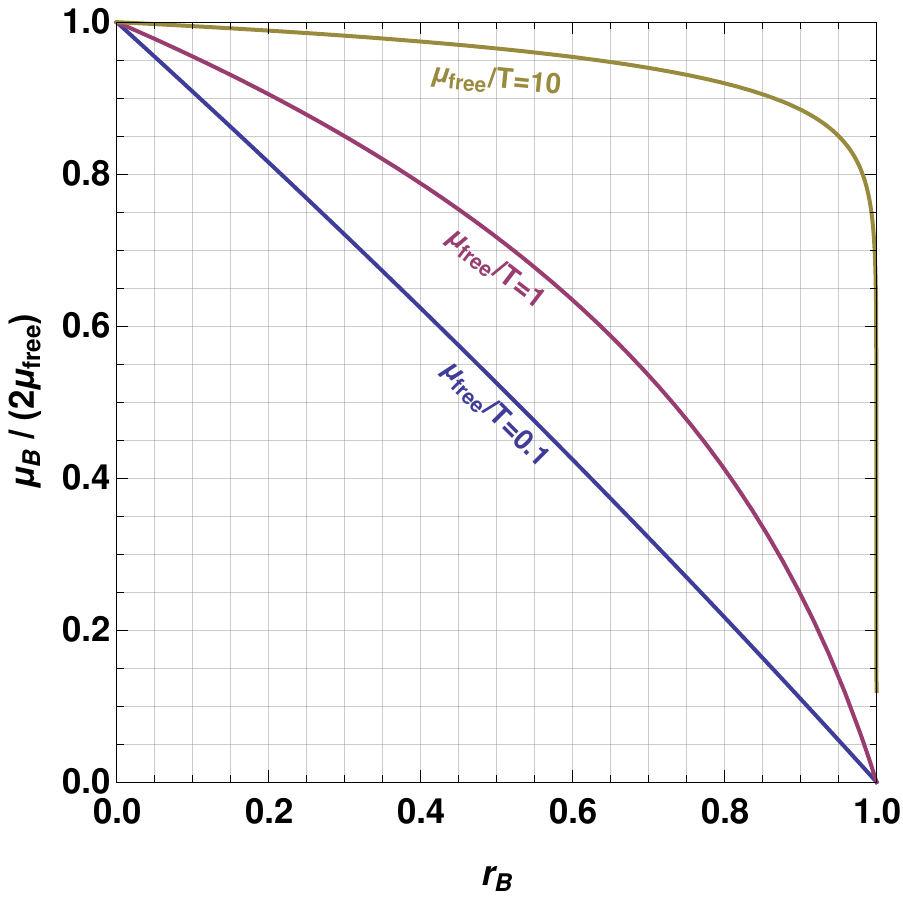}
\caption{The extended Saha equilibrium relation \eqref{eq:SahaEquilibrium_Metastable_ChemicalPot} between the chemical potentials of metastable bound states and free species, $\mu_{\Bcal}$ and $\mu_{\rm free}$ respectively, and $r_{\Bcal}$, given by \cref{eq:EfficiencyFactors} in terms of the bound-state decay, ionisation and transition rates. $r_{\Bcal}=0$ 
reproduces the standard Saha equilibrium for stable bound states, while $r_{\Bcal} \simeq 1$ corresponds to bound states that decay much more rapidly than they get ionised.}
\label{fig:SahaMetastable}
\end{figure}

We discern two limits of \cref{eq:SahaEquilibrium_Metastable,eq:SahaEquilibrium_Metastable_ChemicalPot}.
\begin{itemize}
\item 
At temperatures much larger than the  binding energies, $T \gg |{\cal E}_{\Bcal}|$, the ionisation rates are large, consequently $r_{\Bcal} \ll 1$. Thus $n_{\Bcal}/n_{\Bcal}^\eq \simeq ( n_{\rm free}/n_{\rm free}^{\eq} )^2$ is independent of ${\cal B}$. 
This implies that the bound states are in chemical equilibrium with the free species, $\mu_{\Bcal} = 2\mu_{\rm free}$, as well as among themselves, however the equilibrium is maintained via the scattering-to-bound transitions and does not depend on the bound-to-bound transitions or bound-state decays. 
In terms of the bound-state densities, this equilibrium is similar to that of stable bound states. Nevertheless, the DM depletion due to BSF may be sizeable even during this period of $r_{\Bcal} \ll 1$.

\item
In the limit $r_{\Bcal} \to 1$, typically attained at  $T \ll |{\cal E}_{\Bcal}|$, the bound-state densities formally approach their equilibrium values with zero chemical potential, $n_{\Bcal}/n_{\Bcal}^\eq \simeq 1$ or $\mu_{\Bcal} = 0$. 

However, if the free species have developed a large chemical potential such that 
$1-r_{\Bcal} \gtrsim \exp(-2\mu_{\rm free}/T)$, 
it is possible that the BSF processes prevent $n_{\Bcal}$ from reaching $n_{\Bcal}^{\eq}$.  
Neglecting transitions for simplicity, \cref{eq:SahaEquilibrium_Metastable_Alt} yields 
$n_{\Bcal} \simeq n_{\Bcal}^{\eq} + n_{\rm free}^2 \<\sigma_{\Bcal}^{\BSF} \vrel\> / \Gamma_{\Bcal}^\dec$.
Considering that $Y_{\rm free} = n_{\rm free} / s$ decreases only slowly after DM freeze-out, and taking into account the form of the Coulomb Sommerfeld factor for the cross-sections that defines their asymptotic behaviour at late times, $S = 2\pi\zeta/(1-e^{-2\pi\zeta})$ with $\zeta = \alpha/\vrel$, we estimate that the second contribution to $n_{\Bcal}$ decreases roughly as $x^{-6+1/2}$ or $x^{-6+2/3}\exp[-(a x)^{1/3}]$, with $a \approx 27\pi^2 |{\cal E}_{\Bcal}|/m$ , if the long-range interaction between the incoming free particles in the BSF processes is attractive ($\alpha>0$) or repulsive ($\alpha<0$) respectively. 
In either case, the second term in $n_{\Bcal}$ continues to dominate over the first at low temperatures.

\end{itemize}
\Cref{eq:SahaEquilibrium_Metastable} relates the densities of free and bound species in the entire continuum between the above 
limits. The relation \eqref{eq:SahaEquilibrium_Metastable_ChemicalPot} between the chemical potentials is shown in \cref{fig:SahaMetastable}. 
While \cref{eq:SahaEquilibrium_Metastable,eq:SahaEquilibrium_Metastable_ChemicalPot} were derived in the context of an expanding universe, they
hold in any (homogeneous and isotropic) thermodynamic environment where metastable bound states attain a (quasi-)steady state.

\subsection{Validity of the steady-state approximation \label{sec:BoltzmannEqs_ValidityApprox}}

We now return to the steady-state approximation \eqref{eq:BoltzEqsEff_Assumptions_BoundStateDerivative0} in order to determine its regime of validity. Using the solution \eqref{eq:SahaEquilibrium_Metastable} for the $Y_{\Bcal'}/Y_{\Bcal'}^{\eq}$ ratio, the Boltzmann \cref{eq:BoltzmannEqs_Bound} for the bound level ${\cal B}$ can be re-written after trivial rearrangement of the terms, as
\begin{align}
\dfrac{dY_{\Bcal}}{dx} = -\Lambda x \, \<\Gamma_{\Bcal}^\tot\>    
\( Y_{\Bcal} -  w_{\Bcal} Y_{\Bcal}^{\eq} \) ,
\label{eq:BoltzmannEqs_Bound_AfterSolution}
\end{align}
where
$w_{\Bcal} = w^2 - (w^2-1) 
( \<\Gamma_{\Bcal}^\dec\> + \sum_{\Bcal' \ne \Bcal} r_{\Bcal'} \mathbb{T}_{\Bcal'\Bcal} ) / 
\<\Gamma_{\Bcal}^\tot\> .
$
From the definition \eqref{eq:EfficiencyFactors} of the $r_{\Bcal}$ factors, we find that 
\begin{align}
\<\Gamma_{\Bcal}^{\dec}\> 
= \sum_{\Bcal'} r_{\Bcal'} (\mathbb{\Gamma}^{\tot}-\mathbb{T})_{\Bcal'\Bcal} 
= r_{\Bcal} \<\Gamma^{\tot}_{\Bcal}\> 
- \sum_{\Bcal'\ne \Bcal} r_{\Bcal'} \mathbb{T}_{\Bcal'\Bcal} ,
\end{align}
where in the second step we separated the sum into ${\cal B}'={\cal B}$ and ${\cal B}'\ne{\cal B}$. With this, $w_{\Bcal}$ simplifies to
\begin{align}
w_{\Bcal} =w^2 - (w^2-1) r_{\Bcal} ,
\label{eq:wB_final}
\end{align}
i.e.~the attractor solution of \cref{eq:BoltzmannEqs_Bound_AfterSolution} is indeed \eqref{eq:SahaEquilibrium_Metastable}, as expected by construction.

For the validity of our approximation, we must now require that for the attractor solution \eqref{eq:SahaEquilibrium_Metastable}, the left-hand side of \cref{eq:BoltzmannEqs_Bound_AfterSolution} is much smaller than each term in the right-hand side, i.e. 
$|d (w_{\Bcal} Y_{\Bcal}^{\eq}) / dx | \ll 
\Lambda x \<\Gamma_{\Bcal}^{\tot}\> (w_{\Bcal} Y_{\Bcal}^\eq)$.
This implies
\begin{align}
\<\Gamma_{\Bcal}^{\dec}\>    
+ \<\Gamma_{\Bcal}^{\ion}\>    
+ \<\Gamma_{\Bcal}^{\trans}\>
\gtrsim x H \left| \dfrac{d \ln (w_{\Bcal} Y_{\Bcal}^{\eq}) }{dx} \right|,
\label{eq:EquilApprox_ValidityCond}
\end{align}
where 
$H$ is the Hubble parameter.\footnote{The factor $x$ that appears in the right-hand side of the condition \eqref{eq:EquilApprox_ValidityCond} is often omitted in rough estimations of the (de)coupling time. However, it always appears in more precise semi-analytical treatments~\cite{Gondolo:1990dk}.}

Let us now estimate $d \ln (w_{\Bcal} Y_{\Bcal}^{\eq}) / dx$.  This of course depends on the evolution of $w$, which is determined by the Boltzmann \cref{eq:BoltzmannEqs_Eff}. 
Before DM begins to freeze-out, i.e.~at $x \lesssim 30$, $w \simeq 1$ hence $w_{\Bcal} \simeq 1$ and $d \ln (w_{\Bcal} Y_{\Bcal}^{\eq}) / dx \simeq -2$. 
When freeze-out begins, $w$ starts growing exponentially with $x$, roughly $w \propto x^{-3/2}e^x$ (though bound states slow down this growth since they prolong the decoupling.) We discern two regimes, following the discussion in \cref{sec:BoltzmannEqs_Saha}. 
As long as the temperature is much higher than the binding energies, $T \gg |{\cal E}_{\Bcal}|$, then $r_{\Bcal} \ll 1$ and $w_{\Bcal} \simeq w^2$, hence 
$d\ln (w_{\Bcal} Y_{\Bcal}^{\eq}) /dx \sim  |{\cal E}_{\Bcal}|/m - 3/(2x) \simeq  - 3/(2x)$. 
At $T \lesssim |{\cal E}_{\Bcal}|$, $r_{\Bcal} \to 1$ and we recall the discussion in \cref{sec:BoltzmannEqs_Saha} stemming from \cref{eq:SahaEquilibrium_Metastable_Alt}. Collecting everything, we estimate that
%
%
\begin{align}
\left| \dfrac{d \ln (w_{\Bcal} Y_{\Bcal}^{\eq})}{dx} \right|
\sim  \left\{ 
\begin{array}{lr}
2,\qquad\qquad\qquad\quad~ 
T \gtrsim m/30, 
\\
{\cal O} (x^{-1}), \quad~~
m/30 \gtrsim T \gtrsim |{\cal E}_{\Bcal}|,  
\\
{\cal O} (x^{-1}~\text{or}~x^{-2/3}),\quad~
|{\cal E}_{\Bcal}|\gtrsim T.
\end{array}
\right.
\label{eq:dwB/dx}
\end{align}

At $T \gg |{\cal E}_{\Bcal}|$, the ionisation rates are typically very large, such that the conditions \eqref{eq:EquilApprox_ValidityCond} are easily satisfied.\footnote{
As discussed, the ionisation rates are related to the thermally averaged BSF cross-sections. However, the BSF cross-sections are typically rather suppressed at the large relative velocities of the interacting particles that are typical at high temperatures, due to the small overlap of the scattering-state and bound-state wavefunctions in this regime. Upon thermal averaging, this suppression is counteracted in part by the Bose enhancement due to the low-energy boson radiated in the capture process~\cite{vonHarling:2014kha}.} 
If the capture into a particular bound level, hence also the ionisation rate of that level, happen to be suppressed, for example due to a symmetry, then the bound-to-bound transitions may suffice to populate that level and satisfy the condition \eqref{eq:EquilApprox_ValidityCond}. This includes up-scattering processes, which are not Boltzmann suppressed at high temperatures.

At $T \lesssim |{\cal E}_{\Bcal}|$, the ionisation rates become exponentially suppressed. However, by then typically $\<\Gamma_{\Bcal}^{\dec}\> \gg H$, unless a particular bound state is (nearly) forbidden to decay directly into radiation due to a symmetry. In this case, it may still effectively decay via transitions to other levels, thereby satisfying \eqref{eq:EquilApprox_ValidityCond}. 
This dynamics is very different from that encountered in the case of stable bound states, e.g.~Hydrogen recombination in the early universe, where the steady-state conditions cease to be satisfied at $T \lesssim |{\cal E}_{\Bcal}|$ and the bound-state densities freeze-out.\footnote{Of course, as is well known, the Hydrogen recombination in the early universe is further delayed due to the very small baryon-to-photon ratio.}  

\medskip

Formally, the above considerations imply that the solutions \eqref{eq:SahaEquilibrium_Metastable} are good approximations if the conditions \eqref{eq:EquilApprox_ValidityCond} are satisfied for \emph{all} bound levels, since \eqref{eq:SahaEquilibrium_Metastable} was employed in deriving the Boltzmann \cref{eq:BoltzmannEqs_Bound_AfterSolution} for ${\cal B}$, to account for the densities of all levels ${\cal B}' \neq {\cal B}$. However, if there are bound levels that do not satisfy \eqref{eq:EquilApprox_ValidityCond}, then these levels decouple from the rest of the system. Hence, the densities of all other levels, as well as the depletion of the DM density, must not be affected significantly by them. We shall now show that the prescription for computing the DM density given by  \cref{eq:BoltzmannEqs_Eff,eq:CrossSection_Effective,eq:EfficiencyFactors} can still be safely used, even though the number densities of the decoupled levels may deviate from \eqref{eq:SahaEquilibrium_Metastable}.

We first consider the contribution to the effective DM depletion cross-section \eqref{eq:CrossSection_Effective} due to capture into a level $\tilde{{\cal B}}$ that does not satisfy the condition \eqref{eq:EquilApprox_ValidityCond}. 
Even though incorrectly estimated, this contribution would be limited by
\begin{align}
\<\sigma_{\tilde{\Bcal}}^{\BSF} \vrel\> r_{\tilde{\Bcal}} 
&\leqslant
\<\sigma_{\tilde{\Bcal}}^{\BSF} \vrel\> =
\dfrac{\<\Gamma_{\tilde{\Bcal}}^{\ion}\> n_{\tilde{\Bcal}}^{\eq}}{(n^{\eq})^2}
\nn \\
&< x H \left|\dfrac{d\ln (w_{\tilde{\Bcal}} Y_{\tilde{\Bcal}}^{\eq}) }{dx} \right| 
\dfrac{n_{\tilde{\Bcal}}^{\eq} }{ (n^{\eq})^2 }
\ll  \dfrac{x H}{ n^{\eq} } ,
\end{align}
where we took into account that $r_{\tilde{\Bcal}} \leqslant 1$
and $n_{\tilde{\Bcal}}^{\eq} / n^{\eq} \sim e^{-x} \ll 1$.
The above implies that the estimated effect of capture into the $\tilde{{\cal B}}$ level on the relic density would be indeed negligible, as it should. 

Next we examine the effect of the transitions to and from level $\tilde{\cal B}$ on $\<\sigma^{\eff} \vrel \>$. Using the derivative of $\<\sigma^{\eff} \vrel \>$ with respect to the transition rates that will be computed in \cref{sec:Transitions_General}, \cref{eq:sigmavEff_derivative_Positive}, we find that the change in $\<\sigma^{\eff} \vrel \>$ due to a non-zero $\<\Gamma_{\tilde{\Bcal}}^{\trans} \>$ is 
\begin{align}
\delta \<\sigma^{\eff} \vrel\>
&= \dfrac{n_{\tilde{\Bcal}}^{\eq}}{(n^{\eq})^2}    
\sum_{\Bcal} (r_{\Bcal} - r_{\tilde{\Bcal}})^2 
\delta \mathbb{T}_{\Bcal \tilde{\Bcal}}
\leqslant \dfrac{n_{\tilde{\Bcal}}^{\eq}}{(n^{\eq})^2}    
\sum_{\Bcal} \delta \mathbb{T}_{\Bcal \tilde{\Bcal}}
\nn \\
&= \dfrac{n_{\tilde{\Bcal}}^{\eq}}{(n^{\eq})^2}    
\delta \<\Gamma_{\tilde{\Bcal}}^{\trans} \>
\ll \dfrac{xH}{n^{\eq}} ,
\end{align}
where we employed similar considerations as before. Thus, the estimated effect on the relic density is again negligible.

We conclude overall that the effective method consisting of \cref{eq:BoltzmannEqs_Eff,eq:CrossSection_Effective,eq:EfficiencyFactors} may be used for the computation of the DM density without much concern about whether the steady-state condition always holds. On the other hand, the validity of the generalised Saha \cref{eq:SahaEquilibrium_Metastable} relies on the condition \eqref{eq:EquilApprox_ValidityCond}, which should be ensured for general applications.

\section{Transitions between bound states \label{Sec:Transitions}}

\subsection{No transitions \label{sec:Transitions_NoTrans}}	

If the transitions between bound states are negligible or outright forbidden,
$\< \Gamma_{\Bcal}^\trans \> \ll \< \Gamma_{\Bcal}^\dec \> + \<\Gamma_{\Bcal}^\ion \>$, \cref{eq:EfficiencyFactors} becomes
\begin{align}
r_{\Bcal} \simeq
\dfrac{\<\Gamma^\dec_{\Bcal}\>}{\<\Gamma^\dec_{\Bcal}\> + \<\Gamma^\ion_{\Bcal}\>},
\label{eq:EfficiencyFactors_NoTrans}
\end{align}
where we took into account that since $\mathbb{\Gamma}^\dec$ and $\mathbb{\Gamma}^\ion$ are diagonal, then 
$(\mathbb{\Gamma}^\dec + \mathbb{\Gamma}^\ion)^{-1}_{\Bcal \Bcal'} 
= \delta_{\Bcal\Bcal'} / \(\<\Gamma^\dec_{\Bcal}\> + \<\Gamma^\ion_{\Bcal}\> \)$. 
This result was first derived in ref.~\cite{Ellis:2015vaa} by considering a single bound level, and the accuracy of the approximation with respect to the solution of the coupled Boltzmann equations has been checked in \cite{Cirelli:2016rnw,Baldes:2017gzu,Baldes:2017gzw,Cirelli:2018iax}.

\subsection{Rapid transitions \label{sec:Transitions_RapidTrans}}

We now consider the possibility that the transitions between bound states are very rapid, in particular
$\< \Gamma_{\Bcal}^\trans \> \gg \<\Gamma_{\Bcal}^\dec \> + \<\Gamma_{\Bcal}^\ion \>$. In \ref{App:InverseMatrixM}, we show that in this limit,
\begin{align}
(\mathbb{\Gamma}^\tot - \mathbb{T})^{-1}_{\Bcal\Bcal'} \simeq
\dfrac{Y_{\Bcal}^\eq}
{\sum_{\tilde{\Bcal}} \(\<\Gamma_{\tilde{\Bcal}}^\dec\>+ \<\Gamma_{\tilde{\Bcal}}^\ion\> \) Y_{\tilde{\Bcal}}^\eq } ,
\label{eq:TransitionsInverse_RapidLimit}
\end{align}
i.e.~$(\mathbb{\Gamma}^\tot - \mathbb{T})^{-1}$ is independent of the column index. This renders the efficiency factors $r_{\Bcal}$ of \cref{eq:EfficiencyFactors} independent of ${\cal B}$,
\begin{empheq}[box=\widefbox]{align}
r_{\Bcal} \simeq 
\dfrac
{\sum_{\tilde{\Bcal}} \< \Gamma_{\tilde{\Bcal}}^\dec \> Y_{\tilde{\Bcal}}^\eq }
{\sum_{\tilde{\Bcal}} \(\< \Gamma_{\tilde{\Bcal}}^\dec \> + \<\Gamma_{\tilde{\Bcal}}^\ion\> \) Y_{\tilde{\Bcal}}^\eq } .
\label{eq:EfficiencyFactors_FastTrans}
\end{empheq}
\Cref{eq:EfficiencyFactors_FastTrans} signifies that when the bound states are connected, only the averaged decay and ionisation rates, and consequently the averaged BSF cross-section, are important. This is consistent with and generalises the findings of \cite{Oncala:2021swy}, where transitions between bound states in a two-level system were considered. Looking back at \cref{eq:SahaEquilibrium_Metastable}, it is evident that as long as transitions are rapid, the ratios $Y_{\Bcal}/Y_{\Bcal}^\eq$ become independent of ${\cal B}$. This means that the bound states are in chemical equilibrium with each other.

\Cref{eq:EfficiencyFactors_FastTrans} constitutes another major result of the present work that can be readily employed to study systems with a large number of bound levels where bound-to-bound transitions are rapid.

\subsection{Rapid versus no transitions   \label{sec:Transitions_Comparison}}

Comparing \cref{eq:EfficiencyFactors_NoTrans,eq:EfficiencyFactors_FastTrans}, we find after elementary manipulations that bound-to-bound transitions increase the contribution to the effective cross-section~\eqref{eq:CrossSection_Effective} due to the capture into a bound level ${\cal B}$, if
\begin{align}
\dfrac{\<\Gamma_{\Bcal}^\dec\>}{\<\Gamma_{\Bcal}^\ion\>} < 
\dfrac
{\sum_{\Bcal'\neq \Bcal} \<\Gamma_{\Bcal'}^\dec\> Y_{\Bcal'}^{\eq}}
{\sum_{\Bcal'\neq \Bcal} \<\Gamma_{\Bcal'}^\ion\> Y_{\Bcal'}^{\eq}} .
\label{eq:Comparison_NoTrans_FastTrans}
\end{align}
Clearly, this condition is satisfied by some bound states and not by others. For higher angular momentum states, $\Gamma_{\Bcal}^{\dec}$ tends to be suppressed by higher powers in the coupling, and for excited states more generally $\Gamma_{\Bcal}^{\ion}$ tends to remain significant until later. The condition \eqref{eq:Comparison_NoTrans_FastTrans} thus tends to be more easily satisfied for the excited states, which tend to have lower decay-to-ionisation ratios.

The above may seem to suggest that the net effect of bound-to-bound transitions on the effective cross-section \eqref{eq:CrossSection_Effective} depends on the relative strengths of the various BSF cross-sections and decay rates, since transitions between two levels affect their corresponding $r_{\Bcal}$ factors in opposite directions. However, the overall effect of transitions is in fact to enhance the DM depletion. Indeed, the total BSF contributions to the effective cross-section in the limits of \cref{sec:Transitions_NoTrans,sec:Transitions_RapidTrans} are
\begin{subequations}
\label{eq:sigmaEff}
\label[pluralequation]{eqs:sigmaEff}
\begin{align}
\<\sigma^{\eff} \vrel \>_{\rm no~trans} &\supset 
\dfrac{1}{s (Y^\eq)^2}
\sum_{\Bcal} 
\dfrac{a_{\Bcal} \ b_{\Bcal}}{a_{\Bcal} + b_{\Bcal}}    ,
\label{eq:sigmaEff_NoTrans}
\\
\<\sigma^{\eff} \vrel \>_{\rm rapid~trans} &\supset 
\dfrac{1}{s (Y^\eq)^2}
\dfrac
{\(\sum_{\Bcal} a_{\Bcal} \)  \( \sum_{\Bcal} b_{\Bcal}\)}
{\(\sum_{\Bcal} a_{\Bcal} \) +\( \sum_{\Bcal} b_{\Bcal}\)}    ,
\label{eq:sigmaEff_RapidTrans}
\end{align}
\end{subequations}
where 
$a_{\Bcal} \equiv \<\Gamma_{\Bcal}^{\dec}\> Y_{\Bcal}^{\eq}$ 
and 
$b_{\Bcal} \equiv \<\Gamma_{\Bcal}^{\ion}\> Y_{\Bcal}^{\eq}$.  
With the cross-sections expressed in this form, in \ref{App:Inequality} we show, using the Cauchy–Bunyakovsky-Schwarz inequality, that 
\begin{align}
\<\sigma^{\eff} \vrel \>_{\rm no~trans} \leqslant
\<\sigma^{\eff} \vrel \>_{\rm rapid~trans} .
\label{eq:RapidTrans>NoTrans}
\end{align}
The equality holds when all bound states have the same decay-to-ionisation rate [cf.~condition \eqref{eq:Comparison_NoTrans_FastTrans}], which is of course an unrealistic case.

However, \cref{eq:sigmaEff_NoTrans,eq:sigmaEff_RapidTrans} approach each other in the limits,
$\<\Gamma_{\Bcal}^{\dec}\> \ll \<\Gamma_{\Bcal}^{\ion}\>$ and 
$\<\Gamma_{\Bcal}^{\dec}\> \gg \<\Gamma_{\Bcal}^{\ion}\>$, where they both become
\begin{align}
\<\sigma^{\eff} \vrel\> \supset 
\sum_{\Bcal} \dfrac{\<\Gamma_{\Bcal}^{\dec}\> Y_{\Bcal}^{\eq}}{s(Y^{\eq})^2} .
\label{eq:EffectiveCrossSections_limits}
\end{align}
This reduces to
\begin{subequations}
\label{eq:EffectiveCrossSections_Tlimits}
\label[pluralequation]{eqs:EffectiveCrossSections_Tlimits}
\begin{align}
\<\Gamma_{\Bcal}^{\dec}\> \ll \<\Gamma_{\Bcal}^{\ion}\>:&~~
(4\pi x)^{3/2}
\sum_{\Bcal} \dfrac{g_{\Bcal,\eff} \<\Gamma_{\Bcal}^{\dec}\>}{g_{\eff}^2 \, m^3}
\, e^{|{\cal E}_{\Bcal}|/T} ,
\label{eq:EffectiveCrossSections_Tlarge}
\\
\<\Gamma_{\Bcal}^{\dec}\> \gg \<\Gamma_{\Bcal}^{\ion}\>:&~~
\sum_{\Bcal} \<\sigma_{\Bcal}^{\BSF} \vrel\> ,
\label{eq:EffectiveCrossSections_Tlow}
\end{align}
\end{subequations}
where the total cross-section for capture into a bound level ${\cal B}$, $\<\sigma_{\Bcal}^{\BSF} \vrel\>$ has been defined in \cref{eq:sigmaBSF_tot_def}.
The former is realised at high temperatures, $T \gg |{\cal E}_{\Bcal}|$. The effective cross-section is independent of the actual BSF cross-sections, but grows exponentially until it saturates to the latter at low temperatures, $T \lesssim |{\cal E}_{\Bcal}|$~\cite{Binder:2018znk}. In the passage between these two regimes, bound-to-bound transitions matter. For systems that involve many bound levels, and the binding energies and decay rates span a large range of values, this regime may have a significant duration, thereby rendering bound-to-bound transitions very important. This is particularly evident if one or more bound levels have sizeable BSF cross-sections but vanishing or very suppressed decay rates, typically due to a symmetry. Neglecting transitions, such levels do not contribute to the depletion of the DM abundance, while considering transitions, their contribution may be very significant. Such an example can be found in refs.~\cite{Oncala:2021swy,Oncala:2021tkz}.

\subsection{The effect of bound-to-bound transitions globally \label{sec:Transitions_General}}

While the discussion above offers important insight in the effect of bound-to-bound transitions on the depletion of the DM abundance, it does not unambiguously prove that this is monotonic. To examine the impact of transitions beyond the two limiting cases of \cref{sec:Transitions_NoTrans,sec:Transitions_RapidTrans}, here we consider the variation of the efficiency factors $r_{\Bcal}$ and the effective cross-section $\<\sigma^{\eff}\vrel\>$, with respect to the bound-to-bound transition rates.
All indices below refer to bound states, and no summation over repeated indices is implied.

Starting from \cref{eq:EfficiencyFactors} for $r_{\Bcal}$, 
\begin{align}
\dfrac{dr_{\Bcal}}{d\mathbb{T}_{k\ell}} = 
\sum_{\Bcal'} \<\Gamma_{\Bcal'}^\dec\>
\dfrac{d(\mathbb{\Gamma}^\tot - \mathbb{T})^{-1}_{\Bcal'\Bcal}}{d\mathbb{T}_{k\ell}} .
\label{eq:rB_derivative}
\end{align}
The derivative of a matrix can be expressed in terms of the derivative of its inverse. In particular, 
\begin{align}
&\dfrac{d(\mathbb{\Gamma}^\tot - \mathbb{T})^{-1}_{\Bcal'\Bcal}}{d\mathbb{T}_{k\ell}} =
\nn \\
&= - \sum_{i,j}
(\mathbb{\Gamma}^\tot - \mathbb{T})^{-1}_{\Bcal'i}   
\ \dfrac{d(\mathbb{\Gamma}^\tot - \mathbb{T})_{ij}}{d\mathbb{T}_{k\ell}}
\ (\mathbb{\Gamma}^\tot - \mathbb{T})^{-1}_{j\Bcal}   .
\label{eq:InverseMatricesDerivatives}
\end{align}
Based on the definitions \eqref{eq:Matrices_Rates} of $\mathbb{\Gamma}^{\tot}$ and $\mathbb{T}$, 
\begin{align}
\dfrac{d(\mathbb{\Gamma}^\tot - \mathbb{T})_{ij}}{d\mathbb{T}_{k\ell}}  
&= \dfrac{d}{d\mathbb{T}_{k\ell}} 
\(\delta_{ij} \sum_{n\neq j} \mathbb{T}_{nj} - \mathbb{T}_{ij}\) 
\nn \\
&=\delta_{ij} \sum_{n}
\(\delta_{kn} \delta_{\ell j} 
+ \delta_{kj} \delta_{\ell n} \dfrac{Y_n^\eq}{Y_j^\eq} \) 
(1-\delta_{nj})
\nn \\
&- \(\delta_{ki} \delta_{\ell j}
+ \delta_{kj} \delta_{\ell i} \dfrac{Y_i^\eq}{Y_j^\eq}\)
(1-\delta_{ij})
\nn \\
&=(\delta_{ij} - \delta_{ki}) \delta_{\ell j}
 +(\delta_{ij} - \delta_{\ell i})  \delta_{kj} 
\dfrac{Y_\ell^\eq}{Y_k^\eq} .
\label{eq:GammaTot-T_derivative}
\end{align}
\Cref{eq:InverseMatricesDerivatives,eq:GammaTot-T_derivative} and the symmetry property \eqref{eq:DetailedBalance_GammaTotMinusT_Inverse} of $(\mathbb{\Gamma}^{\tot}-\mathbb{T})^{-1}$ imply
\begin{align}
\dfrac{Y_{\Bcal}^\eq}{Y_\ell^\eq}
\dfrac{d(\mathbb{\Gamma}^\tot - \mathbb{T})^{-1}_{\Bcal'\Bcal}}{d\mathbb{T}_{k\ell}} 
&=
\[(\mathbb{\Gamma}^\tot - \mathbb{T})^{-1}_{\Bcal'k}
 -(\mathbb{\Gamma}^\tot - \mathbb{T})^{-1}_{\Bcal' \ell}\]
\nn \\
&\times 
\[(\mathbb{\Gamma}^\tot - \mathbb{T})^{-1}_{\Bcal\ell}
 -(\mathbb{\Gamma}^\tot - \mathbb{T})^{-1}_{\Bcal k} \] .
\label{eq:GammaTot-T_inverse_derivative}
\end{align}
Note that both sides of \cref{eq:GammaTot-T_inverse_derivative} are symmetric in the interchange $k\leftrightarrow\ell$. With this, and recalling the definition \eqref{eq:EfficiencyFactors} of $r_{\Bcal}$, \cref{eq:rB_derivative} becomes
\begin{align}
\dfrac{Y_{\Bcal}^\eq}{Y_\ell^\eq}
\dfrac{dr_{\Bcal}}{d\mathbb{T}_{k\ell}} = 
-(r_{k}- r_{\ell}) 
\[(\mathbb{\Gamma}^\tot - \mathbb{T})^{-1}_{\Bcal k}
- (\mathbb{\Gamma}^\tot - \mathbb{T})^{-1}_{\Bcal \ell} \] .
\label{eq:rB_derivative_final}
\end{align}
From \cref{eq:rB_derivative_final} we deduce that $dr_{\Bcal} / d\mathbb{T}_{k\ell}$ may be either positive or negative, which is consistent with discussion around the condition \eqref{eq:Comparison_NoTrans_FastTrans}.

Next, we examine the variation of the effective cross-section \eqref{eq:CrossSection_Effective},
\begin{align}
&\dfrac{1}{Y_\ell^\eq} 
\dfrac{d \<\sigma^{\eff} \vrel\>}{d\mathbb{T}_{k\ell}} 
=
\dfrac{1}{s (Y^\eq)^2 \, Y_\ell^\eq}
\sum_{\Bcal} Y_{\Bcal}^\eq \<\Gamma^{\ion}_{\Bcal}\>
\dfrac{dr_{\Bcal}}{d\mathbb{T}_{k\ell}} 
\nn \\
&= - \dfrac{(r_{k}- r_{\ell})}{s (Y^\eq)^2}
\sum_{\Bcal} \<\Gamma^{\ion}_{\Bcal}\>
\[(\mathbb{\Gamma}^\tot - \mathbb{T})^{-1}_{\Bcal k}
- (\mathbb{\Gamma}^\tot - \mathbb{T})^{-1}_{\Bcal \ell} \] .
\label{eq:sigmavEff_derivative}
\end{align}
Recalling \cref{eq:1-rB}, the above becomes
\begin{empheq}[box=\widefbox]{align}
\dfrac{1}{Y_\ell^\eq} 
\dfrac{d \<\sigma^{\eff} \vrel\>}{d\mathbb{T}_{k\ell}} 
= \dfrac{(r_{k} - r_{\ell})^2}{s (Y^\eq)^2} \geqslant 0.
\label{eq:sigmavEff_derivative_Positive}
\end{empheq}
This unambiguously proves that bound-to-bound transitions can only enhance the depletion of the DM abundance.

\section{Toy model \label{Sec:ToyModel}}

While bound-state processes in realistic particle physics models involve considerable complexity that is beyond the scope of this work, in order to illustrate some of our results here we introduce a toy model. 
We consider in particular a three-level system with one free species $\chi$ of mass $m$ and dof $g_{\chi}=1$ and two bound levels \textbf{1} and \textbf{2}, with masses $m_{\one} = 2m-|{\cal E}_{\one}|$, $m_{\two} = 2m-|{\cal E}_{\two}|$ and dof $g_{\one}=g_{\two}=1$; ${\cal E}_{\one} < {\cal E}_{\two} < 0$ are the binding energies. To avoid reference to dimensionful quantities, we define
\begin{subequations}
\begin{gather}
\tau^{\ann} \equiv \lambda (\sigma^{\ann} \vrel), \quad
\tau_{\Bcal}^{\BSF} \equiv \lambda (\sigma_{\Bcal}^{\BSF} \vrel), 
\\
\gamma_{\Bcal}^{\dec} \equiv \dfrac{\lambda \<\Gamma_{\Bcal}^{\dec}\>}{m^3}, \quad
\gamma_{\Bcal}^{\ion} \equiv \dfrac{\lambda \<\Gamma_{\Bcal}^{\ion}\>}{m^3}, \quad
\gamma_{\Bcal \to \Bcal'}^{\trans} \equiv 
\dfrac{\lambda \<\Gamma_{\Bcal \to \Bcal'}^{\trans}\>}{m^3} ,
\\
\varepsilon_{\Bcal} \equiv |{\cal E}_{\Bcal}|/(2m),
\end{gather}
\end{subequations}
for ${\cal B} = \mathbf{1,2}$, where $\lambda$ is defined in \cref{eq:lambda}. 
Clearly, $\varepsilon_{\one} > \varepsilon_{\two} > 0$.
The equilibrium densities of the three levels are
\begin{subequations}
\begin{align}
Y^{\eq} &= \dfrac{90}{(2\pi)^{7/2} g_{*\mathsmaller{S}} } x^{3/2}e^{-x}, \\ 
Y_{\Bcal}^{\eq} &= \dfrac{90}{(2\pi)^{7/2}}  
\dfrac{(1-\varepsilon_{\Bcal})^{3/2}}{g_{*\mathsmaller{S}}}
(2x)^{3/2} e^{-2x}  e^{2x \, \varepsilon_{\Bcal}} .
\end{align}
\end{subequations}
The detailed balance \cref{eq:DetailedBalance_IonBSF,eq:DetailedBalance_B2Btransitions} imply
\begin{subequations}
\begin{align}
\gamma_{\Bcal}^{\ion} &= 
\dfrac{\tau_{\Bcal}^{\BSF}}{(1-\varepsilon_{\Bcal})^{3/2}}
\dfrac{e^{-2x \, \varepsilon_{\Bcal}}}{(4\pi x)^{3/2}} , 
\\
\gamma_{\one \to \two}^{\trans} &= 
\gamma_{\two \to \one}^{\trans} \, 
e^{-2x (\varepsilon_{\one}-\varepsilon_{\two})} .
\end{align}
\end{subequations}
The effective Boltzmann \cref{eq:BoltzmannEqs_Eff} becomes
\begin{align}
\dfrac{dY}{dx} &= -\dfrac{\tau^\eff}{x^2} [Y^2 - (Y^{\eq})^2],
\end{align}
with
\begin{align}
\tau^{\eff} &= \tau^{\ann} + 
\sum_{\Bcal = \one, \two} r_{\Bcal} \tau_{\Bcal}^\BSF.
\end{align}
For our system,
\begin{subequations}
\begin{align}
r_{\one} &= \dfrac
{
\gamma_{\one}^{\dec} + 
\gamma_{\one\to\two}^{\trans}
\(\frac{\gamma_{\two}^{\dec}}
{\gamma_{\two}^{\dec} + \gamma_{\two}^{\ion} + \gamma_{\two\to\one}^{\trans} } \)
}
{\gamma_{\one}^{\dec}
+\gamma_{\one}^{\ion} 
+\gamma_{\one\to\two}^{\trans} 
\(\frac{\gamma_{\two}^{\dec}+\gamma_{\two}^{\ion}}
{\gamma_{\two}^{\dec} + \gamma_{\two}^{\ion} + \gamma_{\two\to\one}^{\trans} } \)
} ,
\\
r_{\two} &= \dfrac
{
\gamma_{\two}^{\dec} + 
\gamma_{\two\to\one}^{\trans}
\(\frac{\gamma_{\one}^{\dec}}
{\gamma_{\one}^{\dec} + \gamma_{\one}^{\ion} + \gamma_{\one\to\two}^{\trans} } \)
}
{\gamma_{\two}^{\dec}
+\gamma_{\two}^{\ion} 
+\gamma_{\two\to\one}^{\trans} 
\(\frac{\gamma_{\one}^{\dec}+\gamma_{\one}^{\ion}}
{\gamma_{\one}^{\dec} + \gamma_{\one}^{\ion} + \gamma_{\one\to\two}^{\trans} } \)
} ,
\end{align}
where we set $1-\varepsilon_{\one} \simeq 1-\varepsilon_{\two} \simeq 1$ for simplicity, since $\varepsilon_{\Bcal} \ll 1$ for weakly bound states. In the limit of very rapid transitions, 
$\gamma_{\two\to\one}^{\trans} \gg \gamma_{\Bcal}^{\dec}, \gamma_{\Bcal}^{\ion}$, the above become
\begin{align}
r_{\one} \simeq r_{\two} \simeq \dfrac
{\gamma_{\one}^{\dec} + \gamma_{\two}^{\dec} e^{-2x(\varepsilon_{\one}-\varepsilon_{\two})} }
{\gamma_{\one}^{\dec}+\gamma_{\one}^{\ion} 
+(\gamma_{\two}^{\dec}+\gamma_{\two}^{\ion}) e^{-2x(\varepsilon_{\one}-\varepsilon_{\two})} 	
} .
\end{align}
\end{subequations}

For the numerical implementation of the model, we fix $g_{*\mathsmaller{S}} = 10^2$ for simplicity. To imitate a system that freezes-out at $x\sim 30$ in the presence of direct annihilations only, we set $\tau^{\ann}=10^{15}$, and pick $\tau_{\two}^{\BSF} =  10 \tau_{\one}^{\BSF}=10^2 \, \tau^{\ann}$ to ensure that BSF is significant with respect to direct annihilations and that capture to excited states is significant with respect to capture to the ground state. We set 
$\varepsilon_{\one} = 10\,\varepsilon_{\two} = 10^{-3}$, 
to ensure two well separated and weakly coupled bound states, and consider three different cases for the bound-state decay rates, summarised in \cref{tab:ToyModels}. Note that the values $\varepsilon \sim 10^{-3}$ and $\gamma^{\dec} \sim 10^{11}$ correspond roughly to those of the ground state of a Coulomb potential with $\alpha \sim 0.1$, for $m\sim \TeV$.
\begin{table}[h]
\centering	
\renewcommand{\arraystretch}{1.3}
\begin{tabular}{|c|c|c|c|}
\hline	
Cases	&$A$	&$B$	&$C$
\\ \hline 
$\gamma_{\one}^{\dec}$ &$10^{11}$ 	&0 			&$10^{11}$	 
\\ \hline
$\gamma_{\two}^{\dec}$ &0 			&$10^{11}$ 	&$10^{11}$ 
\\ \hline
\end{tabular}
\caption{Different versions of the toy model. \label{tab:ToyModels}}
\end{table}

\begin{figure}[h!]
\centering
\includegraphics[width=0.48\textwidth]{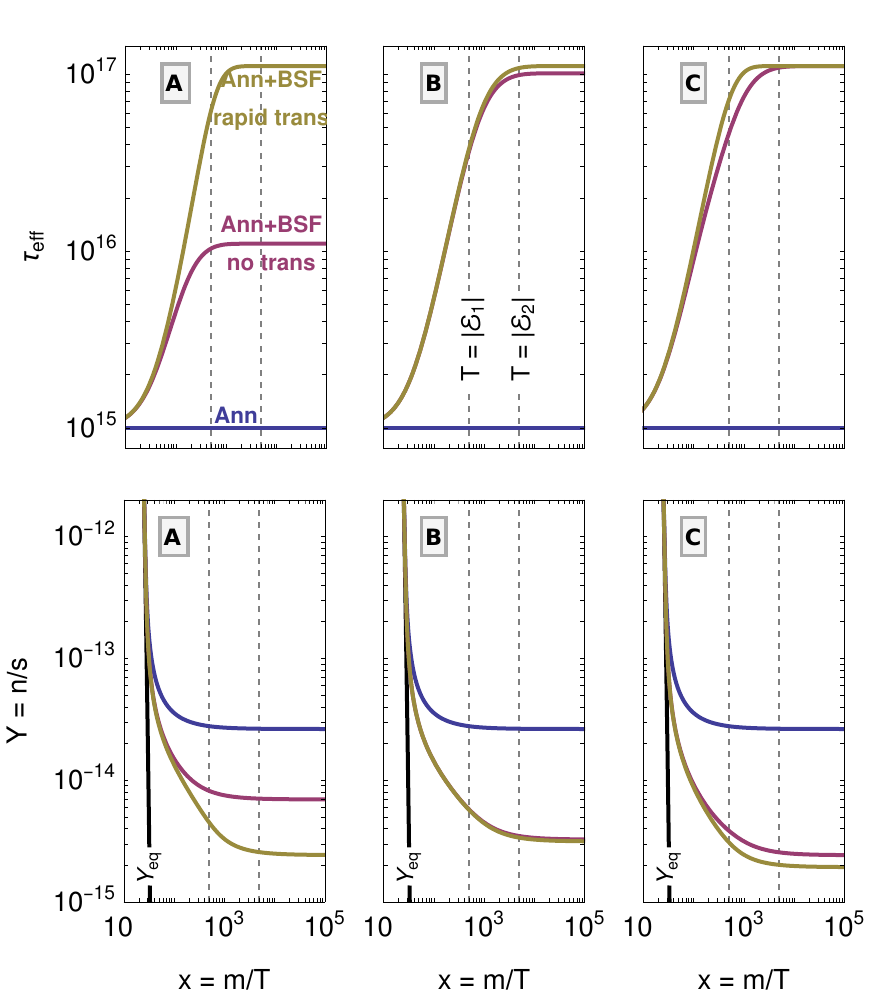}
\caption{The evolution of $\tau_{\eff}$ and the relic abundance for the three models of \cref{tab:ToyModels}. We compare the cases of direct annihilation only (blue), and annihilation plus BSF with no or very rapid bound-to-bound transitions (purple and yellow, respectively).}
\label{fig:ToyModel_TimeEvolution}
\end{figure}

For the models of \cref{tab:ToyModels}, we present the evolution of $\tau_{\eff}$ and the $\chi$ abundance in \cref{fig:ToyModel_TimeEvolution}. Comparing the cases $\gamma_{2\to1} =0$ and $\gamma_{2\to1}$ larger than all other rates, we see that transitions enhance $\tau_{\eff}$ in all cases. The effect is more pronounced when the excited state decays via down-scattering to the ground state (case $A$). When the ground state decays via up-scattering to the excited state, the effect of transitions is small albeit non-zero (case $B$). If the direct decay rates of both bound levels are significant, bound-to-bound transitions have a moderate effect (case $C$). 

\begin{figure}[h!]
\centering
\includegraphics[width=0.48\textwidth]{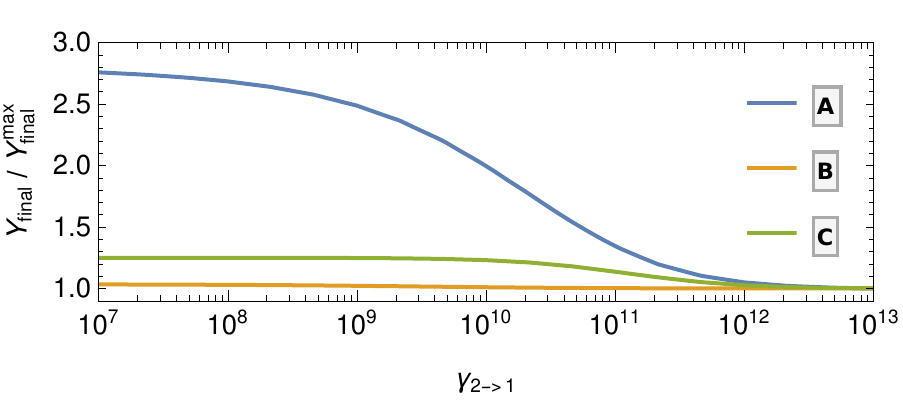}
\caption{
The final abundance $Y_{\rm final}$ as a function of the bound-to-bound transition rate $\gamma_{2\to1}$, normalised to the maximum value $Y_{\rm final}^{\rm max}$ obtained for very large $\gamma_{2\to1}$. The three curves correspond to the models of \cref{tab:ToyModels}.}
\label{fig:ToyModel_Transitions}
\end{figure}

\Cref{fig:ToyModel_Transitions} shows the monotonic decrease of the final density with increasing transition rate, in agreement with the result of \cref{sec:Transitions_General}. We find that $Y_{\rm final}^{\rm no~trans} / Y_{\rm final}^{\rm max} \simeq 2.82, 1.035, 1.25$ for cases $A$, $B$ and $C$ respectively.

\section{Conclusion \label{Sec:Concl}}

The emergence of long-range effects around and above the TeV mass regime within the thermal-relic DM scenario impel a thorough reconsideration of DM freeze-out. It is now well established that the formation and decay of metastable bound states can alter the relic density of stable species even by orders of magnitude, thereby changing the predicted DM mass and couplings to other particles and affecting all experimental signatures. Designing targeted experimental searches for DM and accurately interpreting the experimental results therefore necessitates a robust calculation of the DM freeze-out that will account for all bound-state effects. This is particularly important as experiments are now at the onset of exploring the multi-TeV energy scale.

In this work, we have presented a framework for the computation of the DM relic density when metastable bound states exist in the spectrum of the dark sector. 
This requires in general solving a coupled system of Boltzmann equations that describes the interplay of direct annihilation, bound-state formation, ionisation, transition and decay processes. Here we have proven in full generality that this system can be reduced to one Boltzmann equation of the standard form for DM freeze-out, with the entire dark sector dynamics encoded in a single object, the effective DM depletion cross-section, that we computed in closed form [cf.~\cref{eq:BoltzmannEqs_Eff,eq:CrossSection_Effective,eq:EfficiencyFactors}.]
This result allows to easily incorporate bound-state effects in freeze-out computations, including in public codes.

Of particular interest is the importance of bound-to-bound transitions for the DM depletion via BSF. We have shown that overall, bound-to-bound transitions can only increase the DM depletion rate [\cref{eq:sigmavEff_derivative_Positive}.]
This suggests that excited states may play a more important role, thus DM may be heavier than previously anticipated.
In fact, because the capture into excited states renders higher partial waves relevant for the DM freeze-out~\cite{Baldes:2017gzw}, this result supports the conclusion that partial-wave unitarity does not constrain the mass of thermal relic DM as previously considered~\cite{Baldes:2017gzw}.

We finish by noting that our results generalise the Saha ionisation equilibrium to the case of metastable bound states. We have derived in closed form the relation between the densities of the bound states and the unbound species, in terms of the inherent parameters of the theory under consideration: the bound-state ionisation, transition and decay rates, combined into a single parameter for each bound state, the efficiency factor [cf.~\cref{eq:SahaEquilibrium_Metastable,eq:SahaEquilibrium_Metastable_ChemicalPot}]. This result can be adapted to and employed in studying any system that features metastable bound states.

\appendix
\section*{Appendices}

\section{The limit of rapid transitions  \label{App:InverseMatrixM}}

We want to determine the inverse of the matrix $M\equiv\mathbb{\Gamma}^\tot - \mathbb{T}$, in the regime where the transition rates between bound states are much larger than the decay and ionization rates.  We write $M$ as
\begin{align}
M(\epsilon) = \left[ 
\begin{array}{ccc}
\epsilon_1 + \sum\limits_{i \neq 1} \langle \Gamma^{\text{trans}}_{1\rightarrow i}\rangle 
& \cdots 
&-\langle \Gamma^{\trans}_{N\rightarrow 1} \rangle 
\\
\vdots 
&\ddots 
&\vdots 
\\
-\langle \Gamma^{\trans}_{1\rightarrow N}\rangle
&\cdots 
&\epsilon_N+ \sum\limits_{i \neq N} \langle \Gamma^{\trans}_{N\rightarrow i}\rangle
\end{array}\right].
\label{eq:MatrixM}
\end{align}
where we introduced $\epsilon_i\equiv \<\Gamma^{\dec}_i\>+\<\Gamma_i^{\ion}\>$. 
We will treat these quantities as perturbations and determine the leading order dependence of the determinant of $M$ on $\epsilon_i$.  The inverse of a matrix is simply the adjugate matrix divided by the determinant,
\begin{equation}
M_{ij}^{-1}= \frac{{\rm adj}[M]_{ij}}{\det (M)} \equiv \frac{(-1)^{i+j}m_{ji}}{\det (M)} \ .
\end{equation}
Here $m_{ji}$ is the matrix of minors.
Let us first consider the determinant of $M$. Starting from \cref{eq:MatrixM}, we first replace the $i$-th row with the sum of all rows,
\begin{align}
\text{det}(M)&=
\left| 
\begin{array}{ccc}
M_{1,1}&\cdots&M_{1,N}\\
\vdots& &\vdots\\
M_{i-1,1}&\cdots&M_{i-1,N}\\
\epsilon_1&\cdots&\epsilon_N\\
M_{i+1,1}&\cdots&M_{i+1,N}\\
\vdots& &\vdots\\
M_{N,1}&\cdots&M_{N,N}
\end{array} \right| .
\end{align}
Then, adding to the $j$-th column all other columns, $k \neq j$, each multiplied by $Y_{k}^{\eq}/Y_j^{\eq}$, the $\ell j$ element for $\ell \ne i$, becomes 
$\sum_{k} M_{\ell k} Y_k^\eq/Y_j^\eq 
= \sum_k [\delta_{\ell k} (\epsilon_{k}+\<\Gamma_k^\trans\>) 
- \mathbb{T}_{\ell k}] Y_k^\eq/Y_j^\eq 
=\epsilon_{\ell} Y_\ell^\eq / Y_j^\eq$, where in the last step we used the detailed balance \cref{eq:DetailedBalance_Tmatrix,eq:DetailedBalance_GammaY=TY}. We obtain
\begin{align}
&\text{det}(M) =\left| 
\begin{array}{c|c|c} 
&\epsilon_1 \dfrac{Y_1^\text{eq}}{Y_j^\text{eq}} 
&
\\
M_{ab}
&\vdots
&M_{ab}
\\
 \mathsmaller{a \in [1,i),~b \in [1,j)}
&\epsilon_{i-1} \dfrac{Y_{i-1}^\eq}{Y_j^\eq} 
&\mathsmaller{a \in [1,i),~b \in (j,N]}
\\[1em] \hline 
\epsilon_1 ~\cdots~ \epsilon_{j-1} 
&\dfrac{\sum\limits_{k=1}^N\epsilon_k Y_k^{\eq}}{Y_j^{\eq}} 
&\epsilon_{j+1} ~\cdots~ \epsilon_{N}
\\[1em] \hline 
%
&\epsilon_{i+1} \dfrac{Y_{i+1}^\eq}{Y_j^\eq} 
&
\\
M_{ab}
&\vdots 
&M_{ab}
\\
 \mathsmaller{a \in (i,N],~b \in [1,j)}
&\epsilon_{N} \dfrac{Y_{N}^\eq}{Y_j^\eq}
& \mathsmaller{a \in (i,N],~b \in (j,N]}
\end{array} 
\right| .
\end{align}
The above operations leave the determinant invariant, as well as all entries not contained in the $i$-th row or $j$-th column.
Then, expanding with respect to either the $i$-th row or the $j$-th column, one can notice that to leading order in $\epsilon$ (and up to a sign), the determinant is given by the $(i,j)$-entry times the $(i,j)$-minor of $M$, $m_{ij}(\epsilon)$, evaluated in the limit $\epsilon \to 0$, 
\begin{align}
\text{det}(M(\epsilon))&=
\dfrac{\sum_{k=1}^N\epsilon_k Y_k^{\eq}}{Y_j^{\eq}} 
(-1)^{i+j} m_{ij}(0)
+ \mathcal{O}(\epsilon^2) ,
\label{eq:detM_expansion}
\end{align}
for \emph{any} $1 \leqslant i,j \leqslant N$. Hence, the inverse of $M$ is
\begin{align}
M^{-1}_{ij} &= \dfrac{(-1)^{i+j} m_{ji}}{\text{det}(M)} 
= \dfrac{Y_i^{\eq}}{\sum_{k=1}^N\epsilon_k Y_k^{\eq}} 
+ {\cal O}(\epsilon^0),
\label{eq:Minverse_result}
\end{align}
which implies \cref{eq:TransitionsInverse_RapidLimit}.

\section{Rapid versus no transitions \label{App:Inequality}}

We consider two series $a_n$ and $b_n$, and the expressions 
\begin{align}
S_n \equiv \sum_{j=1}^n
\dfrac{a_j b_j}{a_j+b_j},
\quad
L_n \equiv 
\dfrac
{ \sum_{j=1}^n a_j \sum_{i=1}^n b_i }
{ \sum_{\ell=1}^n (a_\ell+b_\ell)} .
\label{eq:Series_def}
\end{align}
For our purposes, $S_n$ and $L_n$ correspond to \cref{eq:sigmaEff_NoTrans,eq:sigmaEff_RapidTrans} respectively, with 
$a_{\Bcal} \equiv \< \Gamma_{\Bcal}^{\dec} \> Y_{\Bcal}^{\eq}$ and
$b_{\Bcal} \equiv \< \Gamma_{\Bcal}^{\ion} \> Y_{\Bcal}^{\eq}$. 
The difference between the two expressions is
\begin{align}
&L_n - S_n 
= \dfrac
{ \sum_{j,i} a_j b_i }
{ \sum_\ell (a_\ell+b_\ell)} 
-\dfrac
{ \sum_j a_j b_j \prod_{k\neq j} \(a_k + b_k \) }
{ \prod_\ell \(a_\ell + b_\ell\)} 
\nn \\[1ex]
&=\dfrac{
  \sum\limits_{j,i} a_j b_i \prod\limits_k \(a_k + b_k \) 
- \sum\limits_j a_j b_j \prod\limits_{k\neq j} \(a_k + b_k \) \sum\limits_i (a_i+b_i)}
{\sum_\ell (a_\ell+b_\ell) \prod_\ell \(a_\ell + b_\ell \)}  
\nn \\ 
&= \frac{ \sum_{j,i} a_j b_i (a_j+b_j) \prod_{k \neq j} (a_k + b_k)}
{\sum_\ell \(a_\ell +b_\ell \) \prod_\ell \( a_\ell + b_\ell \)}
\nn \\ 
&- \frac{\sum_j a_j b_j \prod_{k \neq j} (a_k +b_k ) \sum_i (a_i + b_i) }
{\sum_\ell \(a_\ell +b_\ell \) \prod_\ell \( a_\ell + b_\ell \)}
\nn \\
&= 
\dfrac{\sum_j a_j \prod_{k\neq j} \(a_k + b_k \) 
\sum_{i}\[ b_i (a_j + b_j) - b_j (a_i + b_i) \]}
{\sum_\ell (a_\ell+b_\ell) \prod_\ell \(a_\ell + b_\ell \)}  
\nn \\ 
&=\sum_j \dfrac{a_j}{a_j + b_j} 
\dfrac{ \sum_i \(b_i a_j - b_j a_i \)}{\sum_\ell (a_\ell+b_\ell) } .
\end{align} 
Rearranging further the above, we obtain
\begin{align}
&L_n - S_n = \sum_j \dfrac{1}{a_j + b_j} \times
\nn \\
&\times 
\(\dfrac{ \sum_i a_j \( a_j (b_i + a_i) -a_ja_i\)-  a_i \( a_j(a_j+b_j)-a_j^2 \)}
{\sum_\ell (a_\ell+b_\ell) } \)
\nn \\ 
&= \sum_j \(\dfrac{a_j^2}{a_j+b_j}\)\(1-\dfrac{\sum_i a_i}{\sum_\ell \(a_\ell + b_\ell \)}\) 
\nn \\
&- \dfrac{\sum_i a_i}{\sum_\ell \(a_\ell + b_\ell \)} 
\( \sum_j a_j-\sum_j \dfrac{a_j^2}{a_j+b_j}\) 
\nn \\ 
&= \sum_j \(\dfrac{a_j^2}{a_j+b_j}\) 
-\dfrac{\( \sum_j a_j\)^2}{\sum_j \(a_j + b_j \)} .
\end{align}
Defining 
$u_j \equiv a_j/\sqrt{a_j^2+b_j^2}$ and 
$v_j \equiv \sqrt{a_j^2+b_j^2}$, this becomes
\begin{align}
L_n - S_n = \sum_j u_j^2 - \dfrac{ \(\sum_{j} u_j v_j \)^2}{\sum_j v_j^2} ,
\end{align}
and by virtue of the Cauchy–Bunyakovsky-Schwarz inequality, we find
\begin{align}
L_n - S_n \geqslant 0.
\end{align}

\section*{Acknowledgements}

K.P. thanks Christiana Vasilaki for useful discussions and input on the validity of the steady-state approximation.  
A.F. and K.P. thank the University of Crete and the FORTH Institute of Astrophysics for hospitality. This work was supported by the following:
\begin{itemize}
\item 
European Union’s Horizon 2020 research and innovation programme under grant agreement No 101002846, ERC CoG ``CosmoChart" (work of K.P.)
\item 
ANR ACHN 2015 grant ``TheIntricateDark" (work of K.P.)
\item 
NWO Vidi grant ``Self-interacting asymmetric dark matter" (work of A.F.~and K.P.)
\item
World Premier International Research Center Initiative (WPI), MEXT, Japan (work of T.B.~and G.W.) 
\item 
JSPS Core-to-Core Program Grant Number JPJSCCA20200002 (work of T.B.)
\item 
JSPS KAKENHI Grant Number 20H01895 and JP21H05452 (work of T.B.)
\end{itemize}

\bibliography{Bibliography.bib}

\end{document}